\documentclass{aa}  

\usepackage{graphicx}
\usepackage{txfonts}
\usepackage{natbib}
\bibpunct{(}{)}{;}{a}{}{,} 

\begin{document}

\title{Star formation in the hosts of GHz peaked spectrum and compact steep spectrum radio galaxies}

\author{
A. Labiano\inst{1,2}
\and
C. P. O'Dea\inst{3}
\and
P. D. Barthel\inst{2}
\and
W. H. de Vries\inst{4,5}
\and
S.A. Baum\inst{6}
}
\offprints{Alvaro Labiano:\\ {\tt labiano at damir.iem.csic.es} \smallskip}

\institute{
Departamento de Astrof\'isica Molecular e Infrarroja, Instituto de Estructura de la Materia (CSIC), Madrid, Spain
\and
Kapteyn Astronomical Institute, Groningen, 9700 AV, The Netherlands 
\and
Department of Physics, Rochester Institute of Technology, Rochester, NY, 14623, USA
\and
University of California, Davis, CA 95616, USA
\and
Lawrence Livermore National Laboratory, IGPP, Livermore, CA 94550, USA
\and
Center for Imaging Science, Rochester Institute of Technology,  Rochester, NY 14623. USA 
}

\date{  }

\abstract
{}
{We searched for star formation regions in the hosts of potentially young radio
galaxies (gigahertz peaked spectrum and compact steep spectrum sources).}
{We used near-UV imaging with the Hubble Space Telescope Advanced Camera for Surveys.}
{We find near-UV light could be the product of recent star formation in
five of the nine observed sources, though other explanations are not currently
ruled out. An additional two sources show marginal detections.
The UV luminosities of the GPS and CSS sources are similar to those
of a sample of nearby large-scale radio galaxies. Stellar-population synthesis models
are consistent
with a burst of recent star formation occurring before the formation of the radio
source. However, observations at other wavelengths and colors are needed
to definitively establish the nature of the observed UV light.
In the CSS source \object{1443+77}, the near-UV
light is aligned with and is co-spatial with the radio source.
We suggest that the UV light  in this source is produced by star formation
triggered and/or enhanced by the radio source.}
{}
 
\keywords{Galaxies: active -- Galaxies: starburst -- Galaxies: evolution -- Galaxies: stellar content -- Galaxies: interaction -- Ultraviolet: galaxies}

\maketitle
%

\section{Introduction}

The relationship between black hole mass and galaxy mass implies that the growth and evolution of black holes (therefore AGN) and their host galaxies must somehow be related \citep[e.g.,][]{Gebhardt00}. Mergers and strong interactions can trigger AGN activity in a galaxy \citep[e.g.,][]{Heckman86, Baum92, Israel98}. These events can also produce instabilities in the ISM and trigger star formation \citep[e.g.,][]{Ho05}. Numerical simulations and models \citep[e.g.,][]{Mellema02,Rees89} suggest that the advancement of the jets through the host galaxy environment can also trigger star formation.
Imaging studies in the ultraviolet (UV) light of large 3CR sources find evidence for
episodes of star formation starting around the time when the radio source was
triggered \citep[i.e. $\lesssim 10^7 - 10^8$~yr,][]{Koekemoer99, Allen02, Chiaberge02, O'Dea01, O'Dea03, Martel02}, suggesting a possible link.  

Gigahertz peaked spectrum (GPS) and compact steep spectrum (CSS) radio sources are
apparently young, smaller \citep[GPS $\lesssim 1$ kpc, CSS $\lesssim$15 kpc,
for a review see][]{O'Dea98} versions of the large powerful radio sources, so they are expected to exhibit signs of more recent star formation. In addition, their subgalactic size makes them excellent probes of the interactions between the expanding lobes and the host. They have not completely broken through the host ISM, so these interactions are expected to be more important than in the larger sources.

Near UV observations are very sensitive to the presence of hot young stars and as such will trace recent star formation events. We have therefore obtained high-resolution HST/ACS near-UV images of these young compact sources to study the morphology and the extent of recent star formation. 

This is the first time a sample of GPS and CSS sources has been imaged in the near-UV. It is also the first time that the relative sizes of radio sources in well-matched samples are used to study time evolution of merger-induced and jet-induced star formation.

\section{Observations and data reduction}

We have obtained high resolution near-UV images with the High Resolution Channel
(HRC) of the Advanced Camera for surveys (ACS) on board the Hubble Space
Telescope, through the F330W filter,  with integration times of 1800 seconds.
The objects observed are GPS and CSS galaxies \object{1117+146}, \object{1233+418}, \object{1345+125}, \object{1443+77}, \object{1607+268}, \object{1814--637}, \object{1934--638}, \object{1946+708}, \object{2352+495} (Table \ref{zlambda}). 
Our sample is chosen to be representative of GPS and CSS sources with z$\lesssim$ 0.5, and nearby enough to eliminate strong effects due to evolution with cosmic time. The objects are drawn primarily from the well-defined samples of \citet{Fanti90}, \citet{Fanti01}, \citet{Stanghellini91} and \citet{Stanghellini97}. The comparison sample of large 3CR sources consists of FR I and FR II sources with redshifts less than 0.1 observed in the near-UV by \citet{Allen02}.

The standard ACS reduction pipeline was used to remove detector signatures such as bias, dark current, flat field and to perform flux calibration. Each target was observed in a three-point dither pattern. The frames were combined with Multidrizzle \citep{Koekemoer02} to correct for geometric distortions and cosmic rays. The resulting ACS images have a signal to noise of $\sim$100 for the brightest objects and $\sim$5 for the faintest.

\begin{figure*}[h]
\centering
\includegraphics[width=0.5\columnwidth]{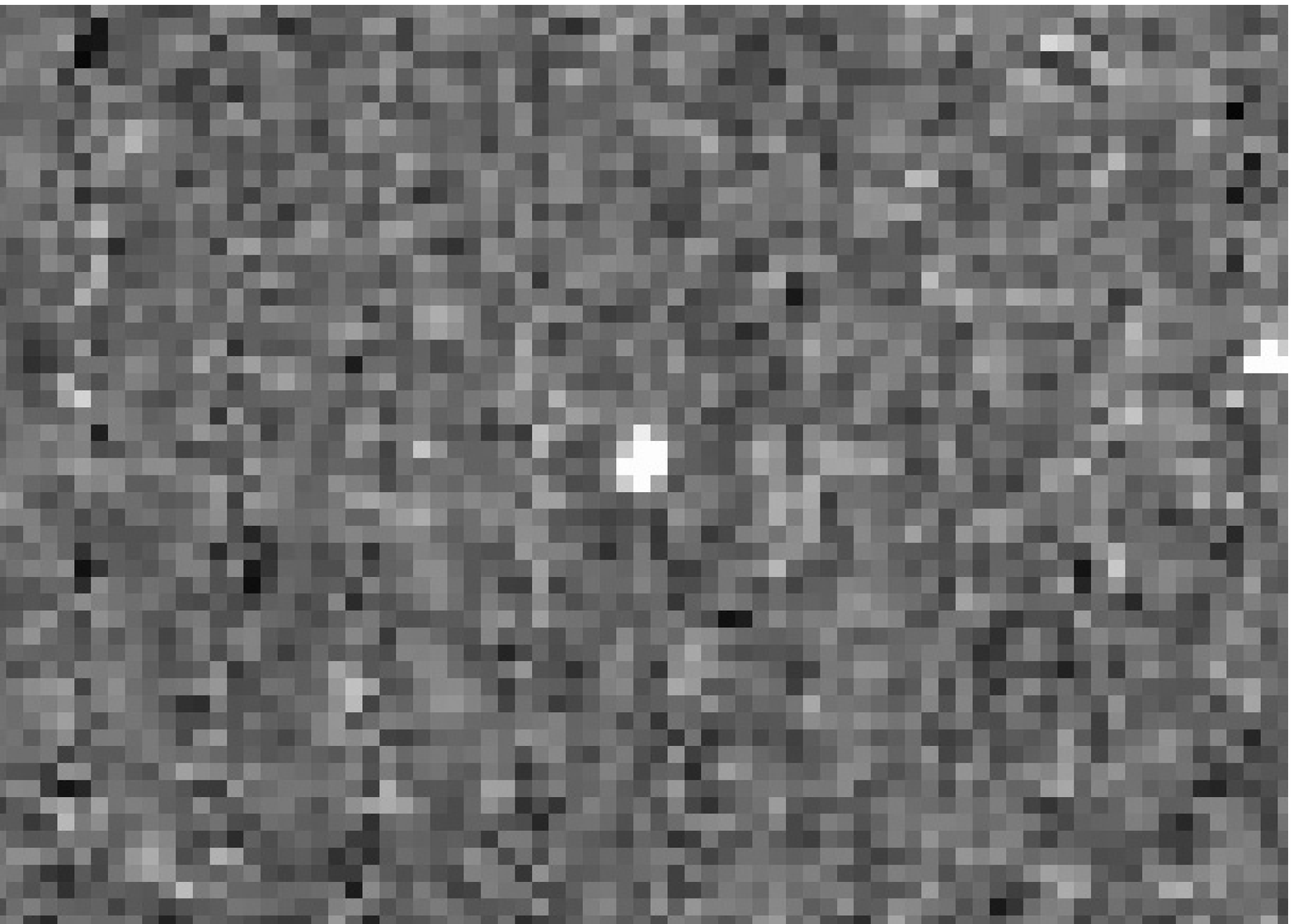} \hfil \includegraphics[width=0.5\columnwidth]{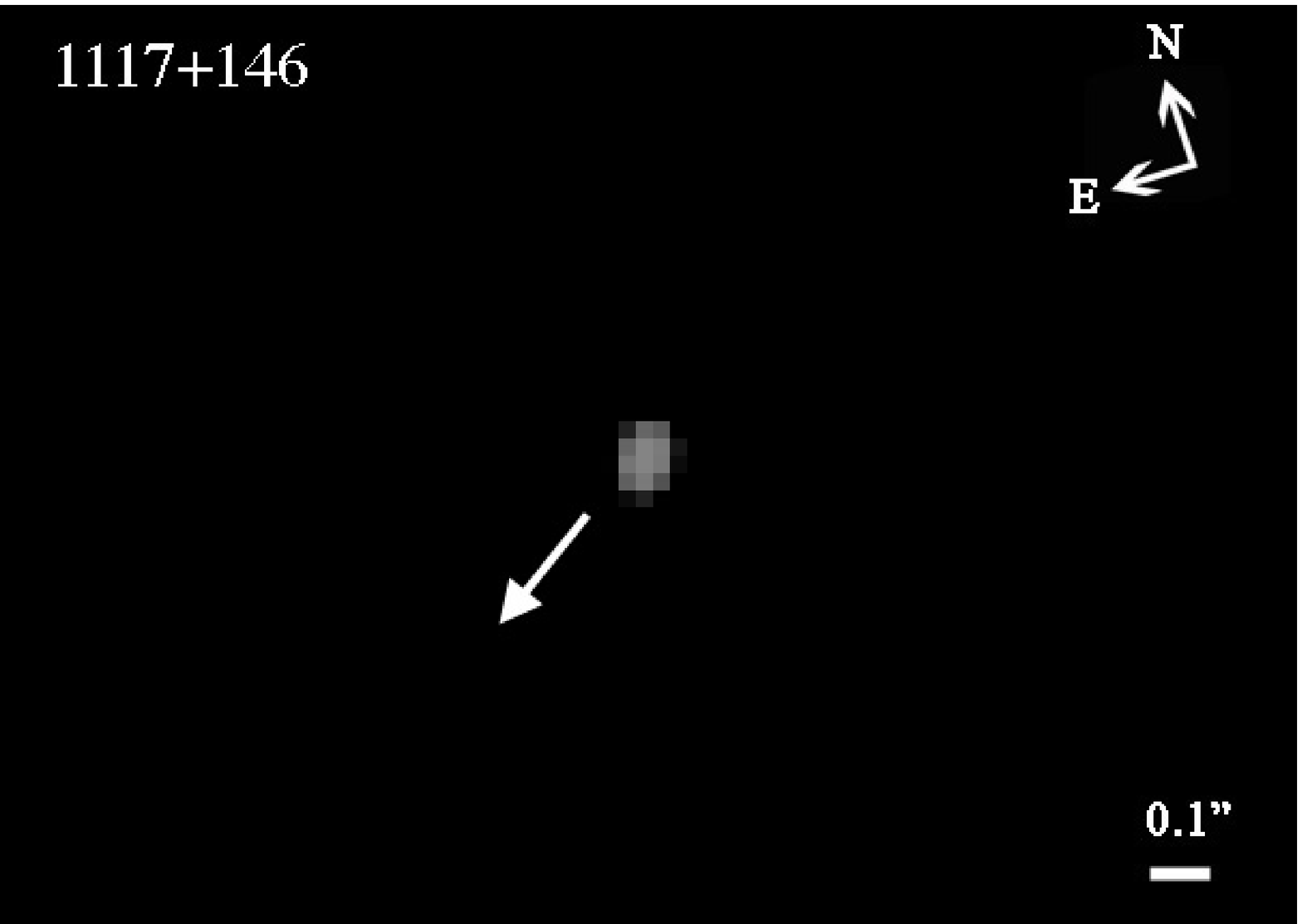} \hfil \includegraphics[width=0.5\columnwidth]{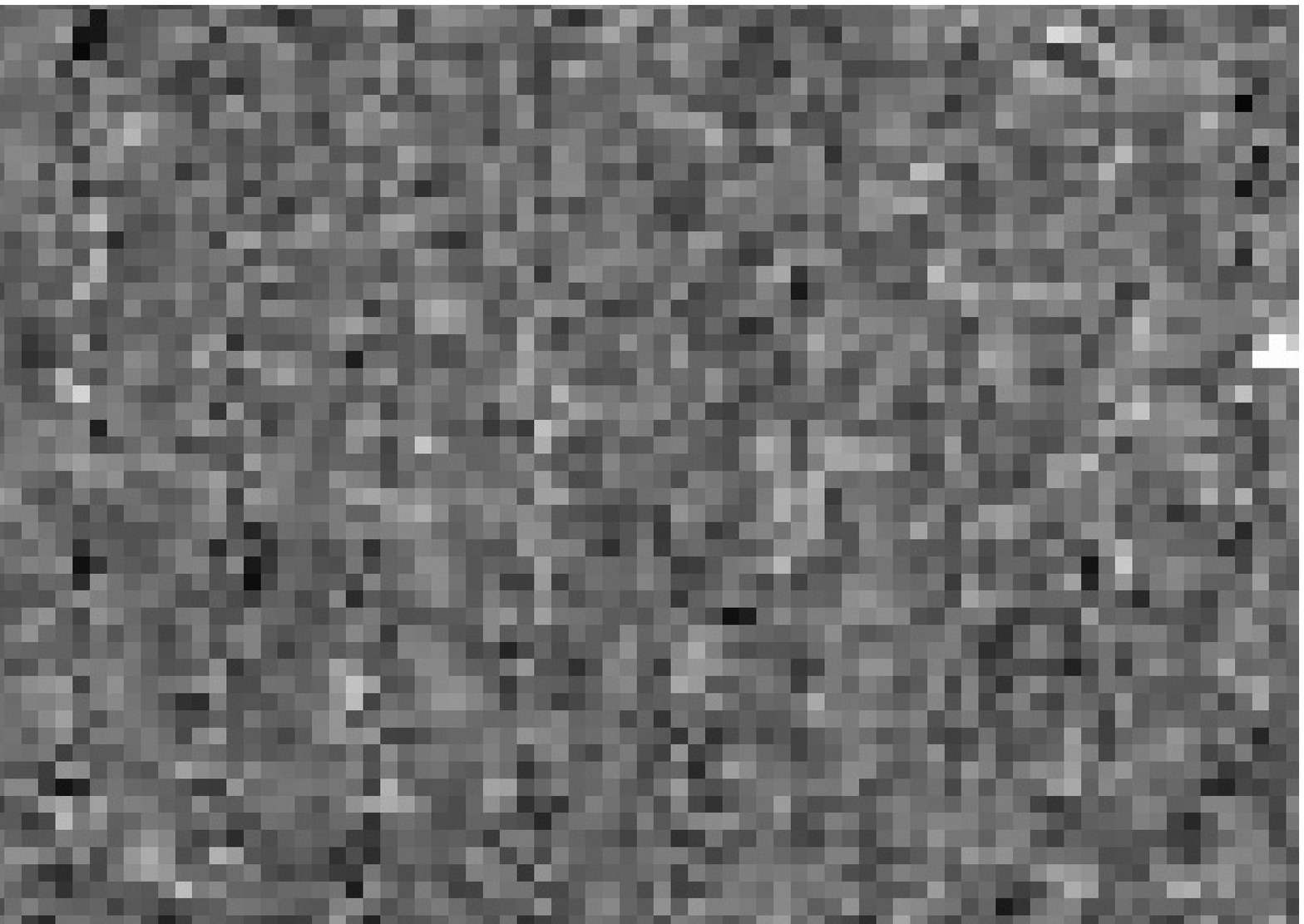} 

\includegraphics[width=0.5\columnwidth]{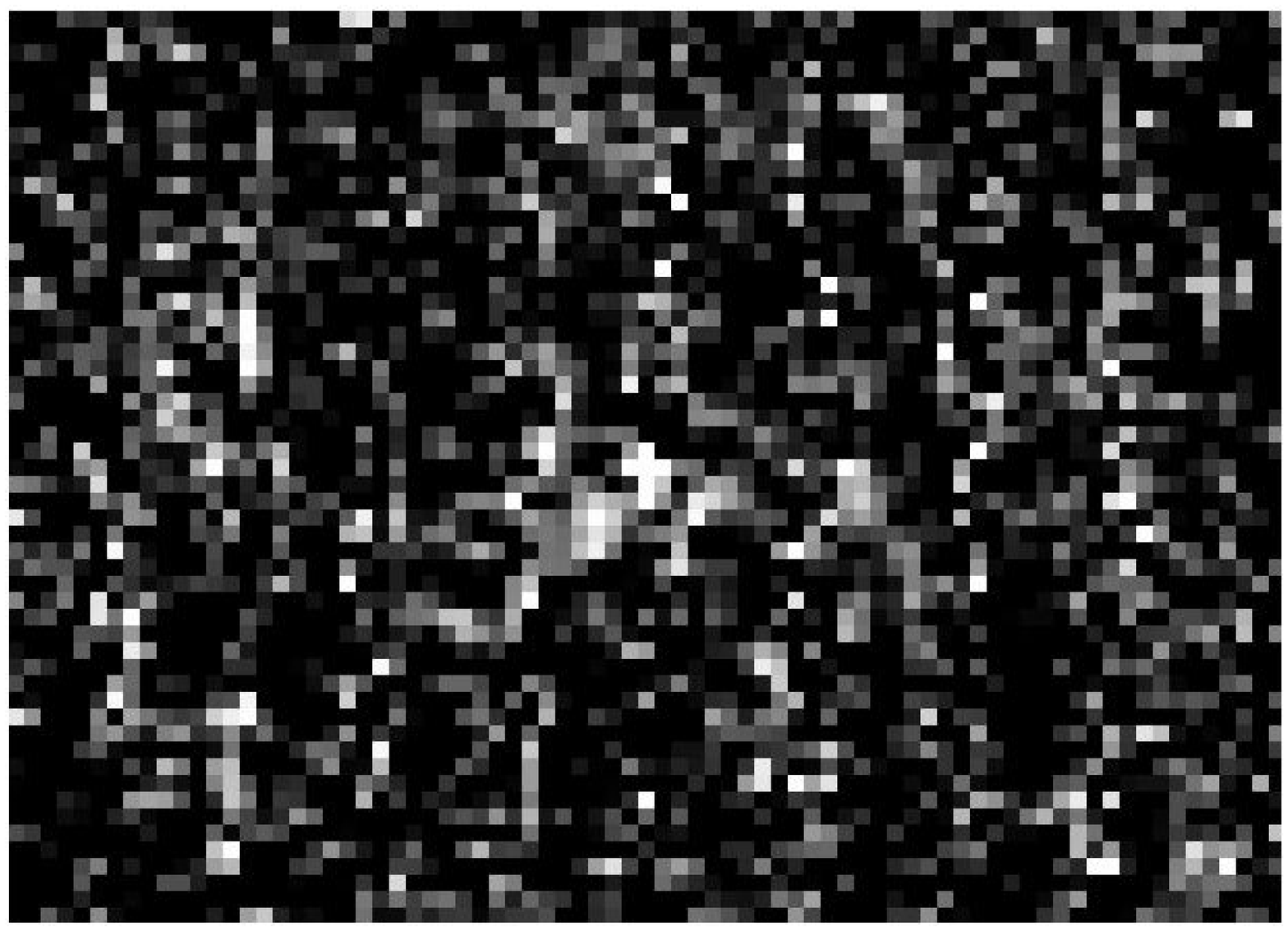} \hfil \includegraphics[width=0.5\columnwidth]{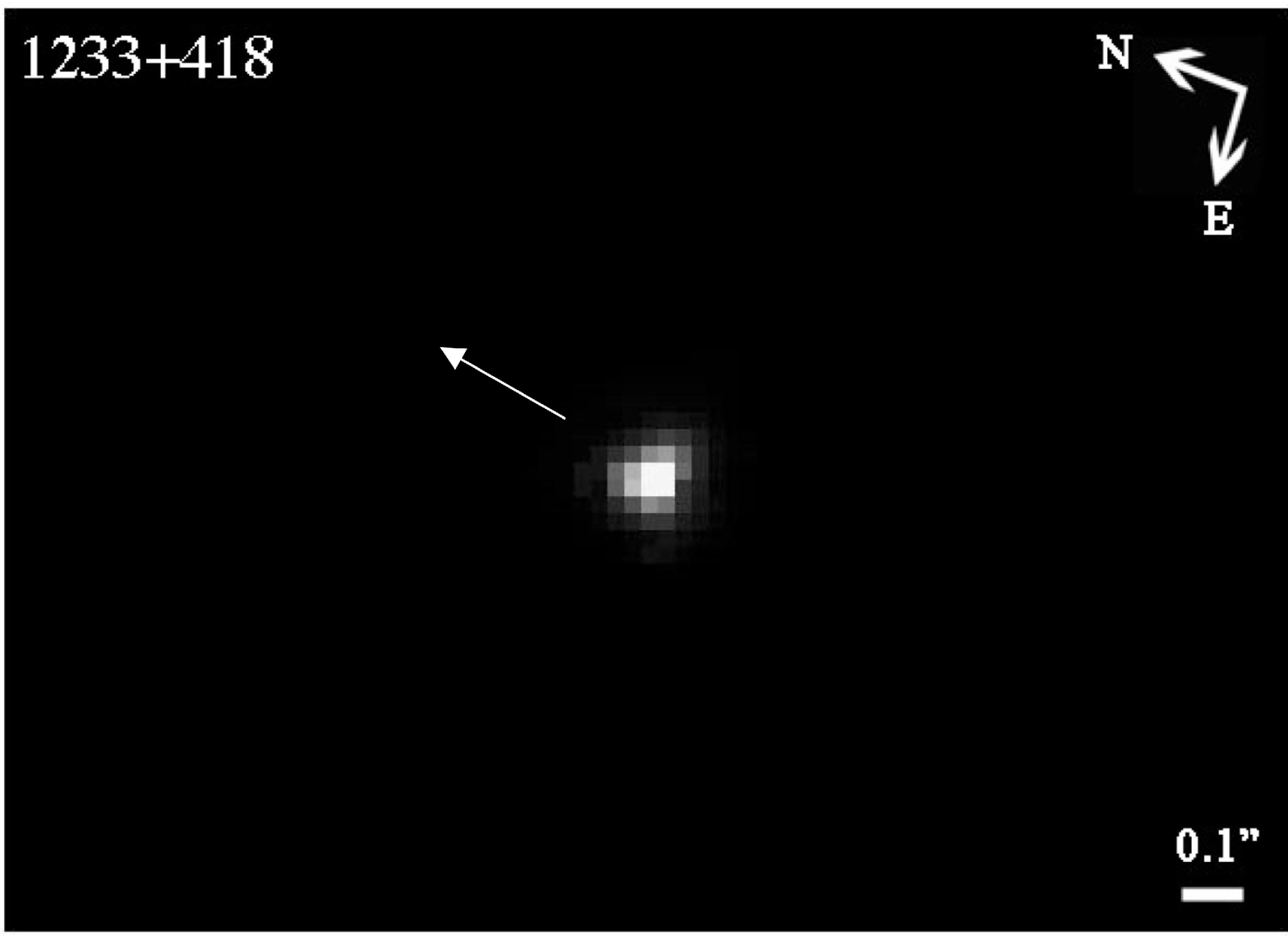} \hfil \includegraphics[width=0.5\columnwidth]{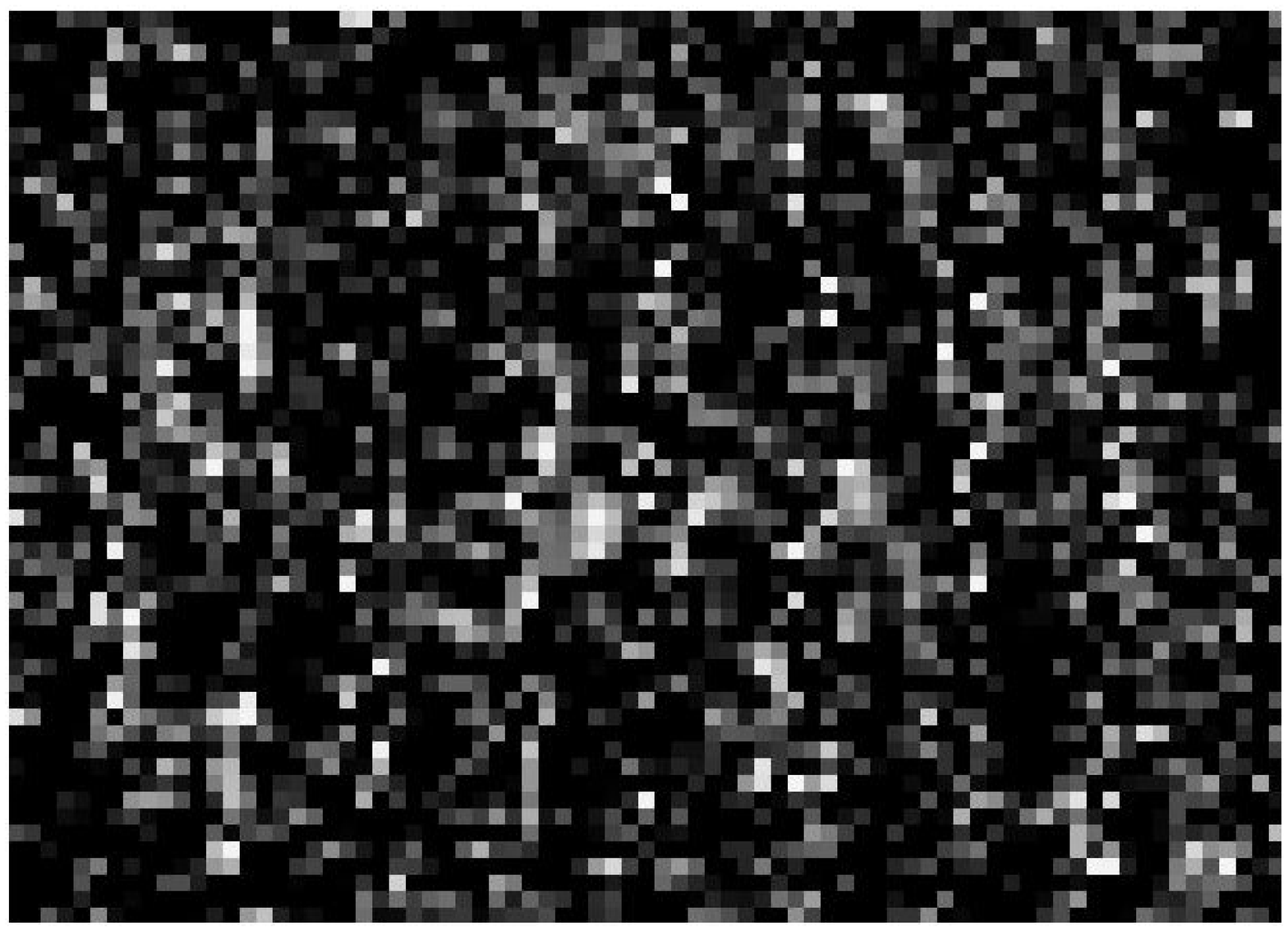} 

\includegraphics[width=0.5\columnwidth]{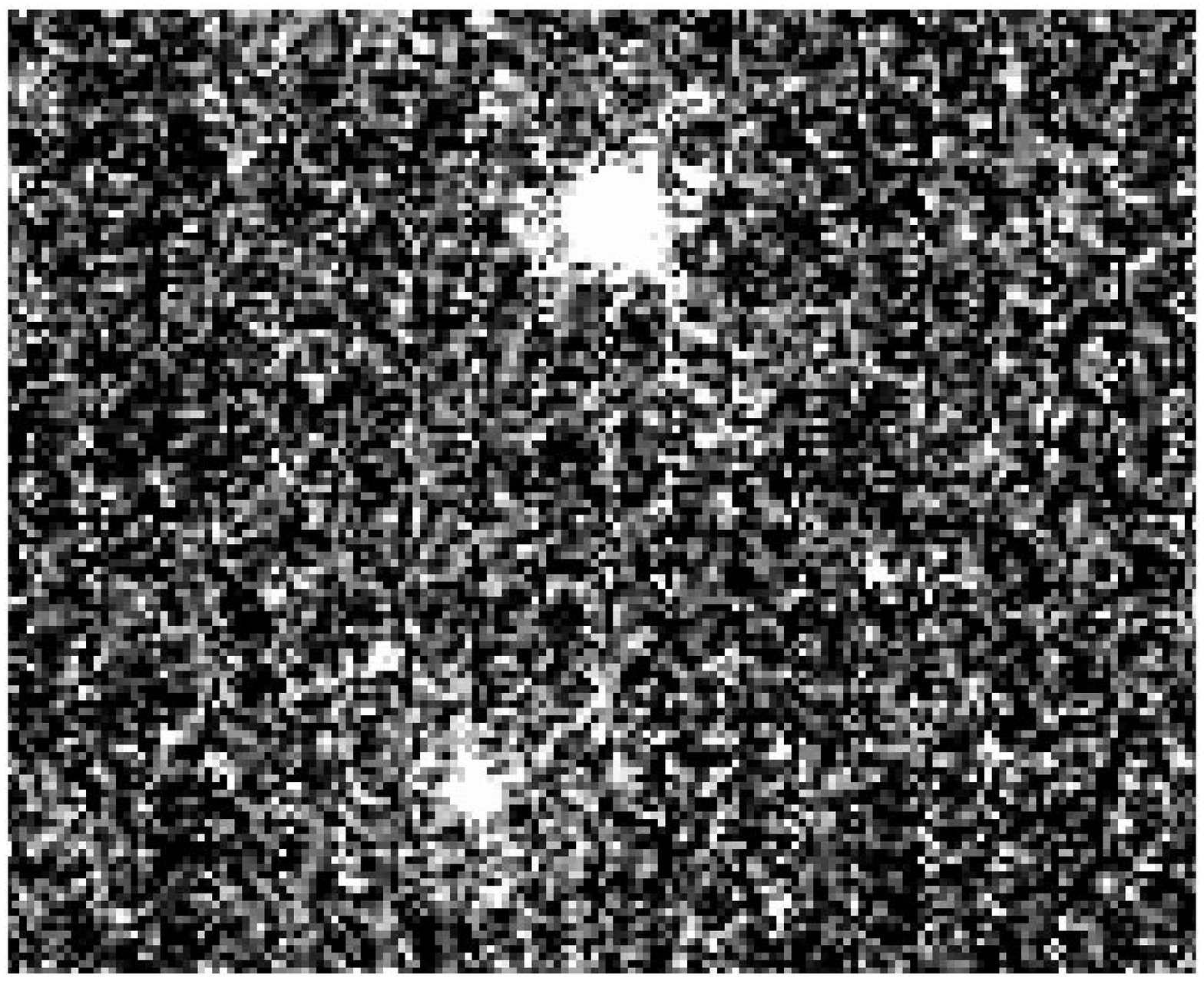} \hfil \includegraphics[width=0.5\columnwidth]{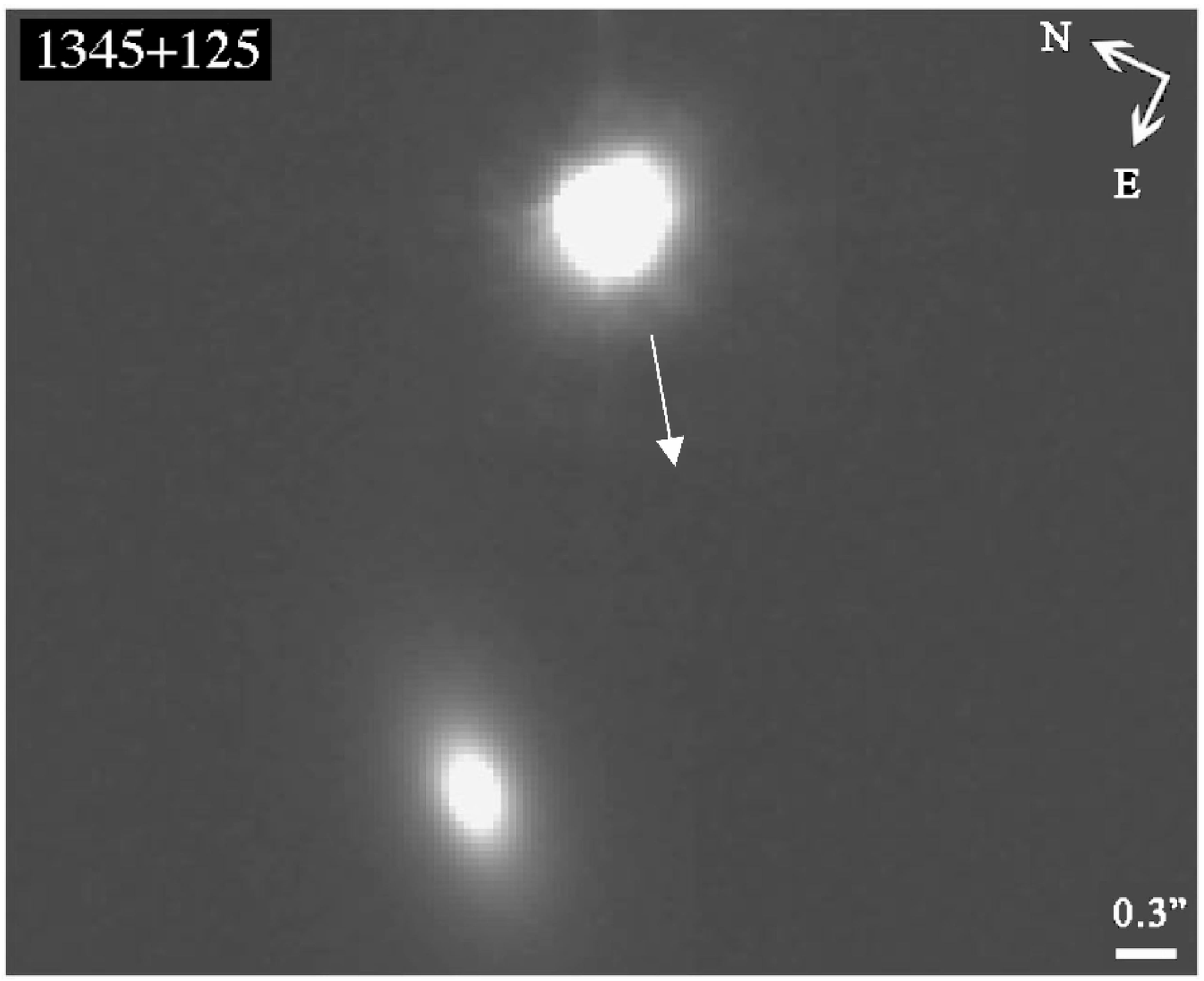} \hfil \includegraphics[width=0.5\columnwidth]{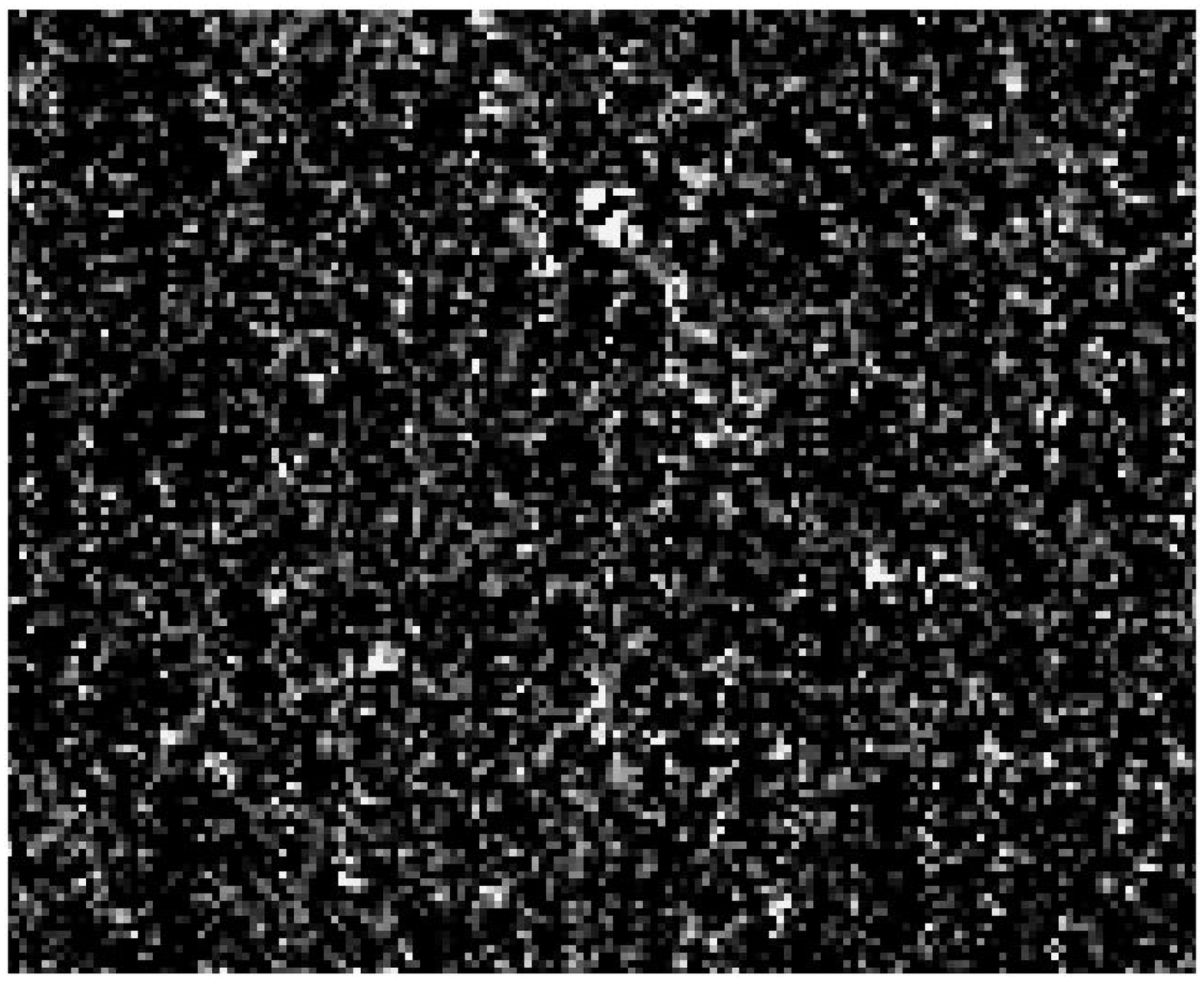} 

\includegraphics[width=0.5\columnwidth]{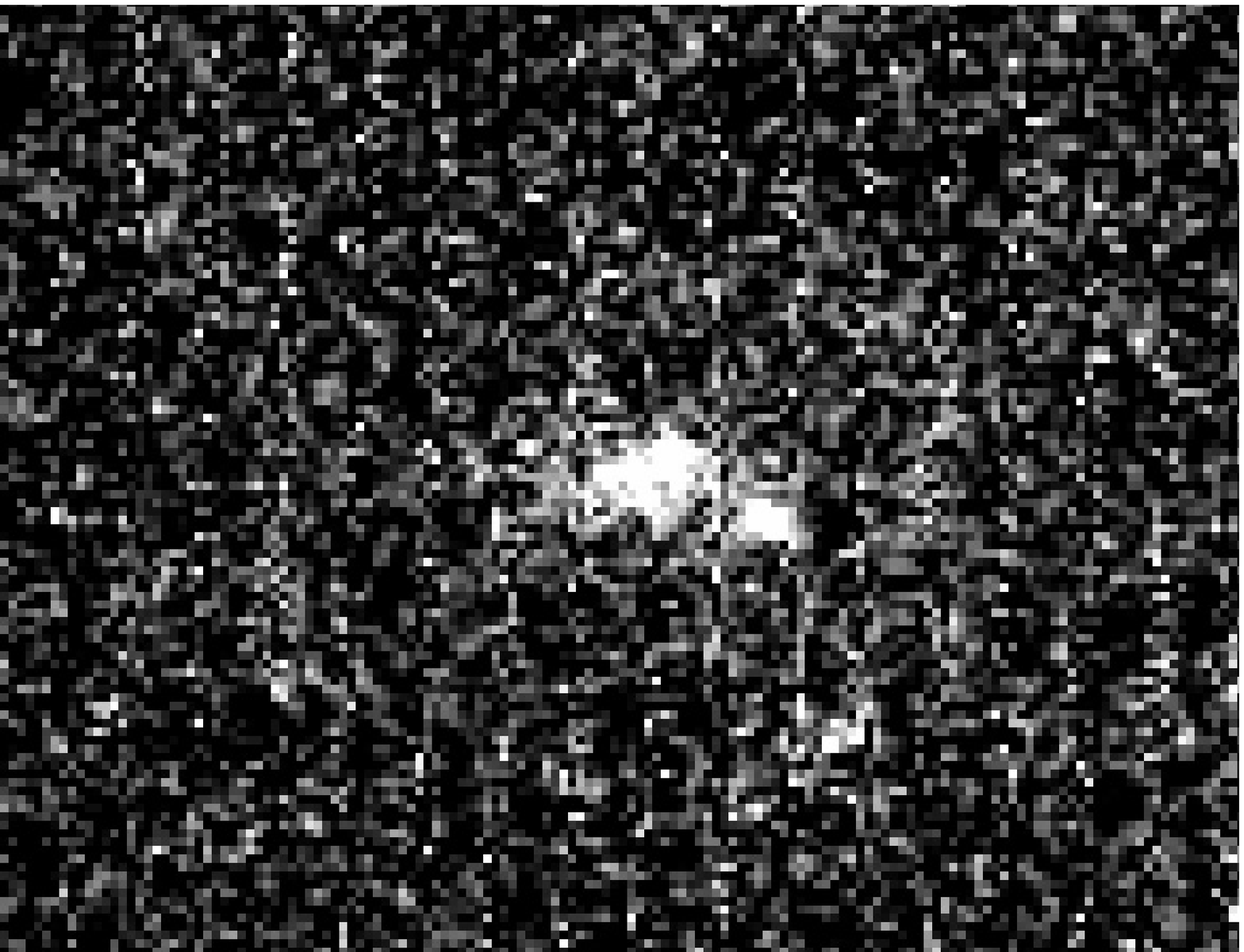} \hfil \includegraphics[width=0.5\columnwidth]{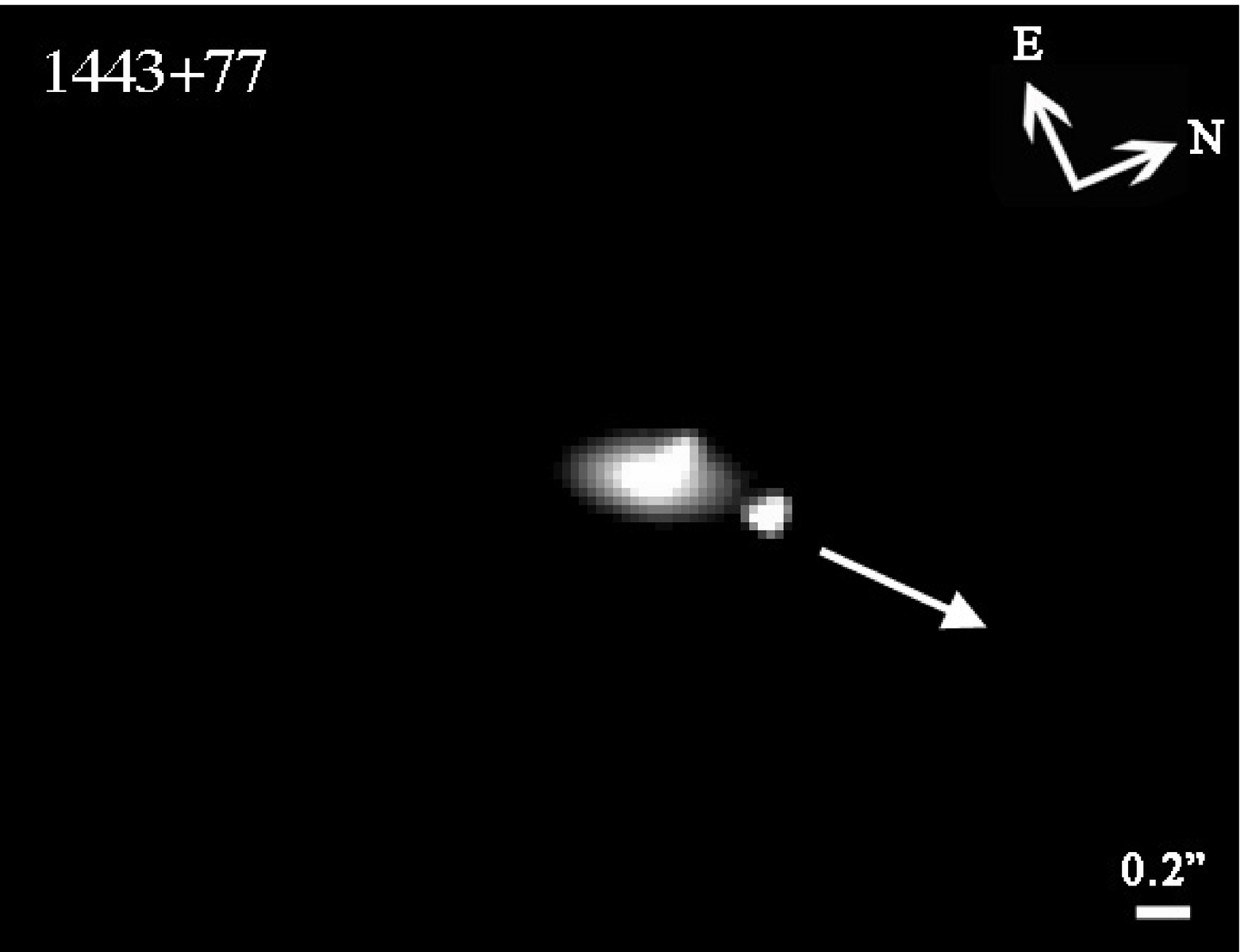} \hfil \includegraphics[width=0.5\columnwidth]{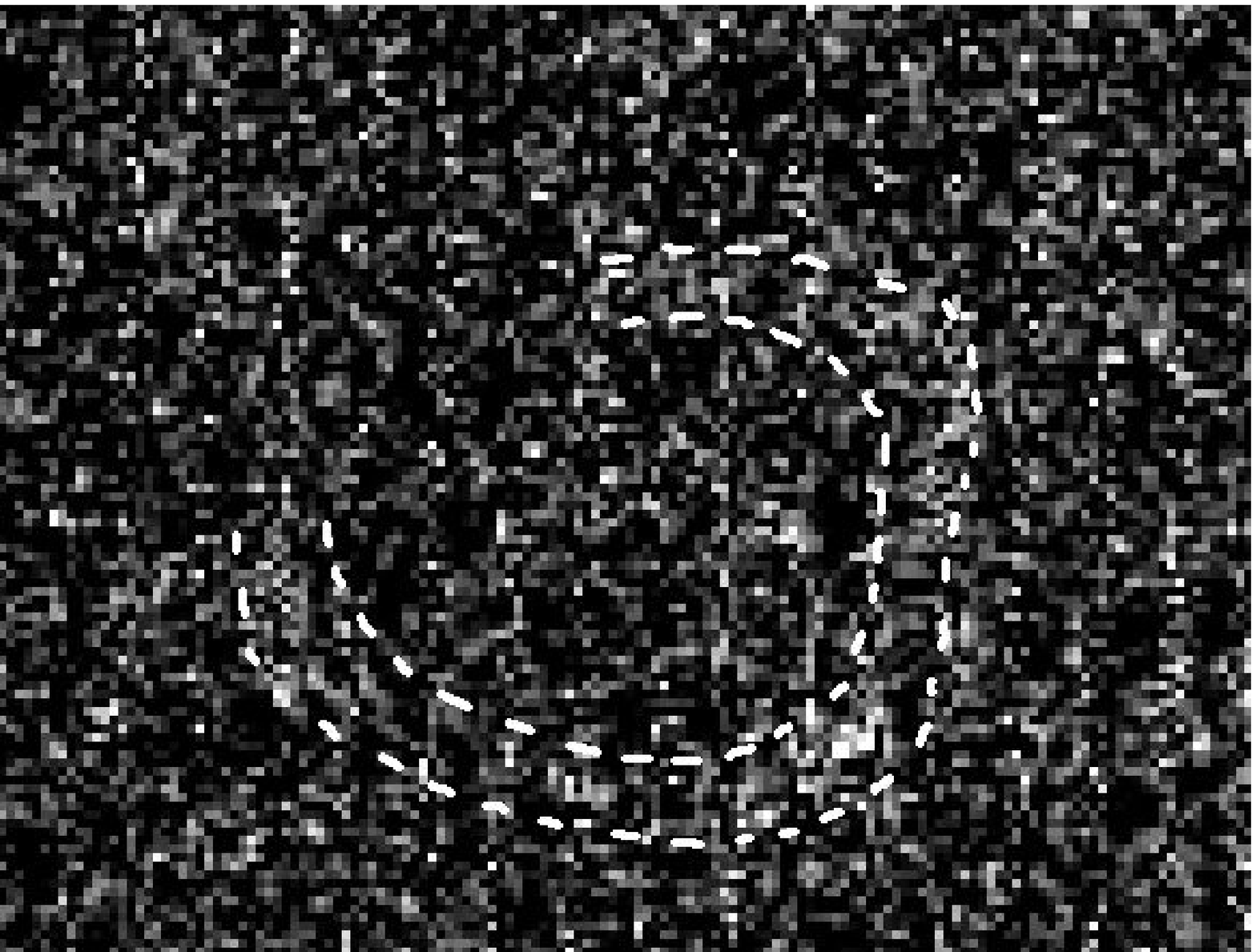} 

\includegraphics[width=0.5\columnwidth]{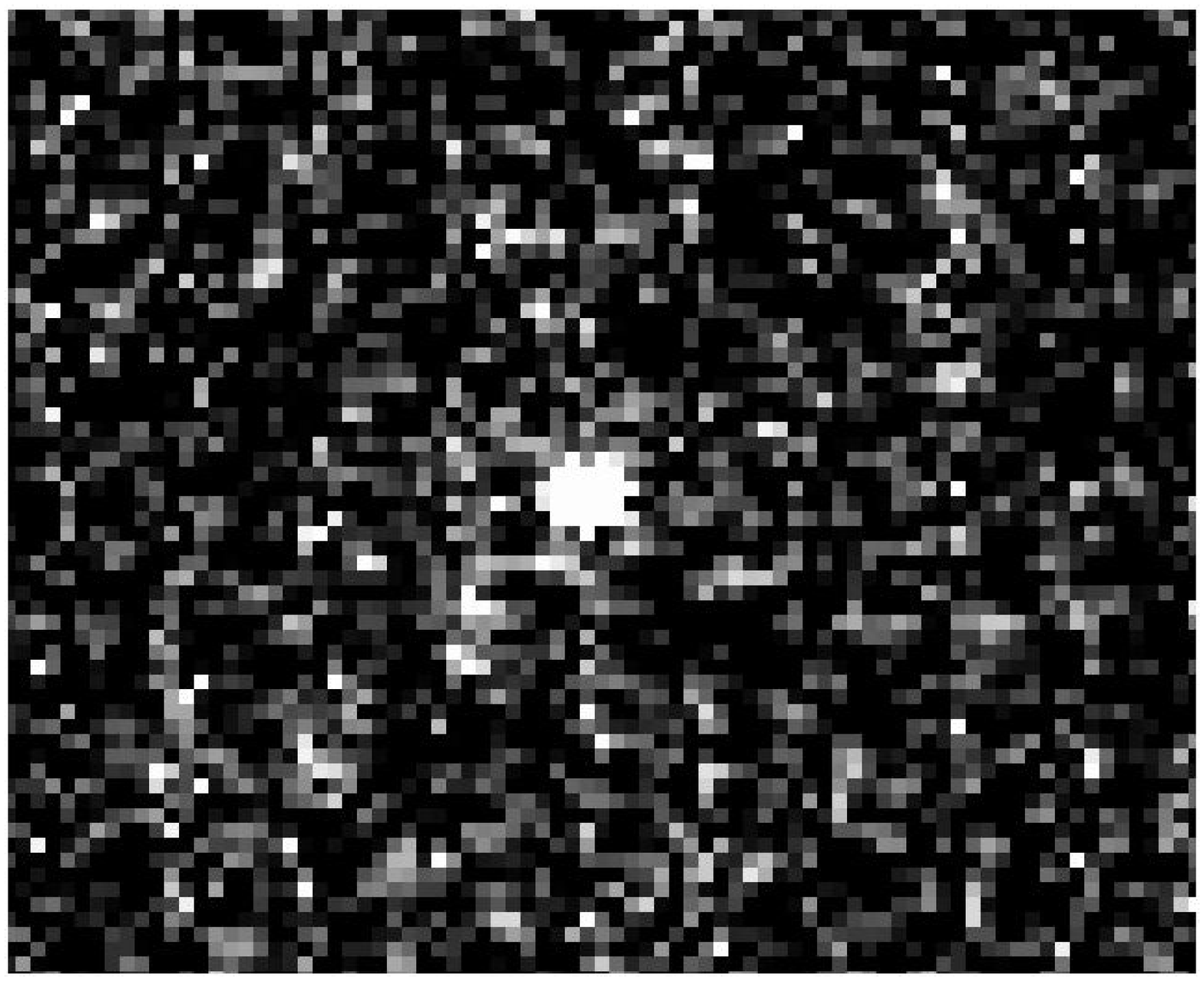} \hfil \includegraphics[width=0.5\columnwidth]{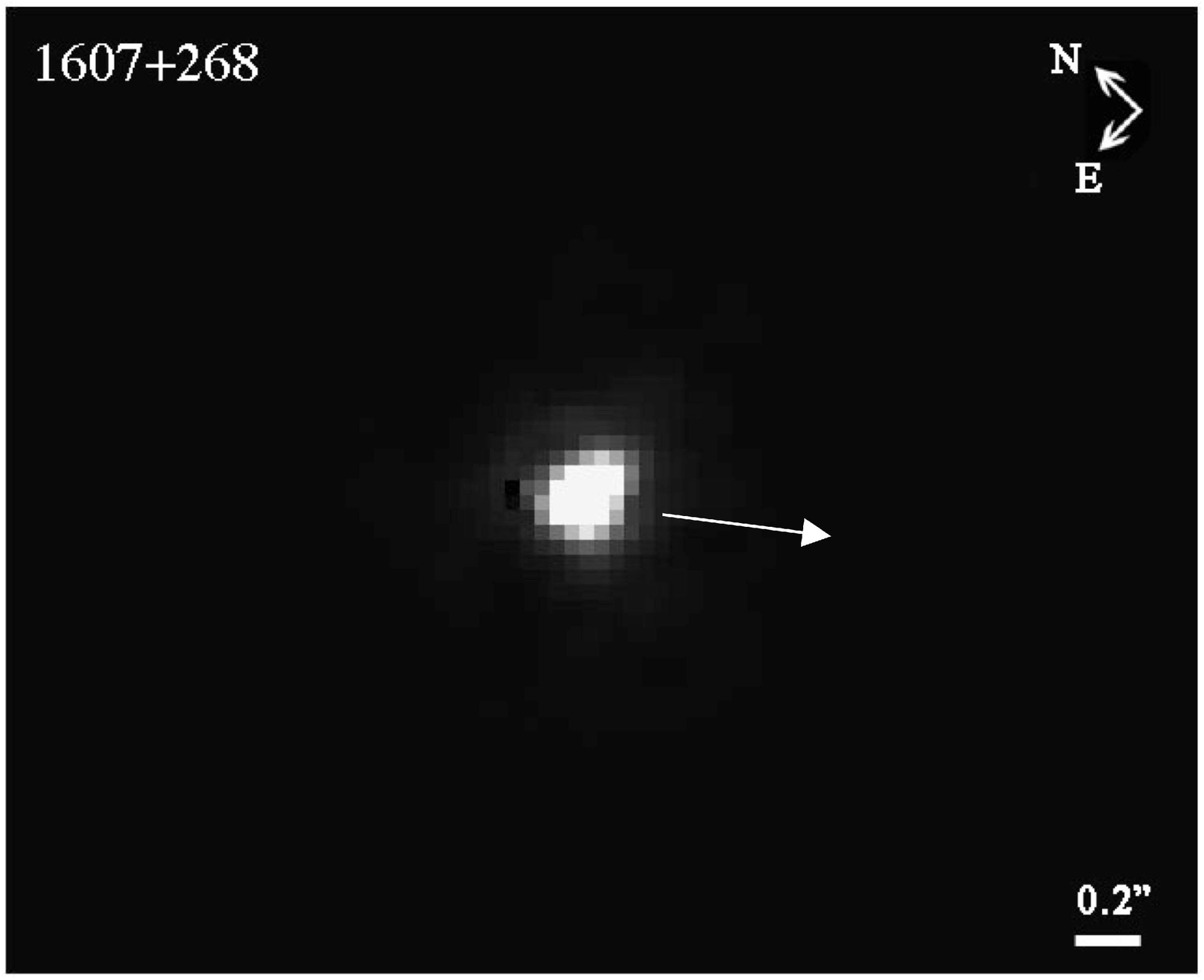} \hfil \includegraphics[width=0.5\columnwidth]{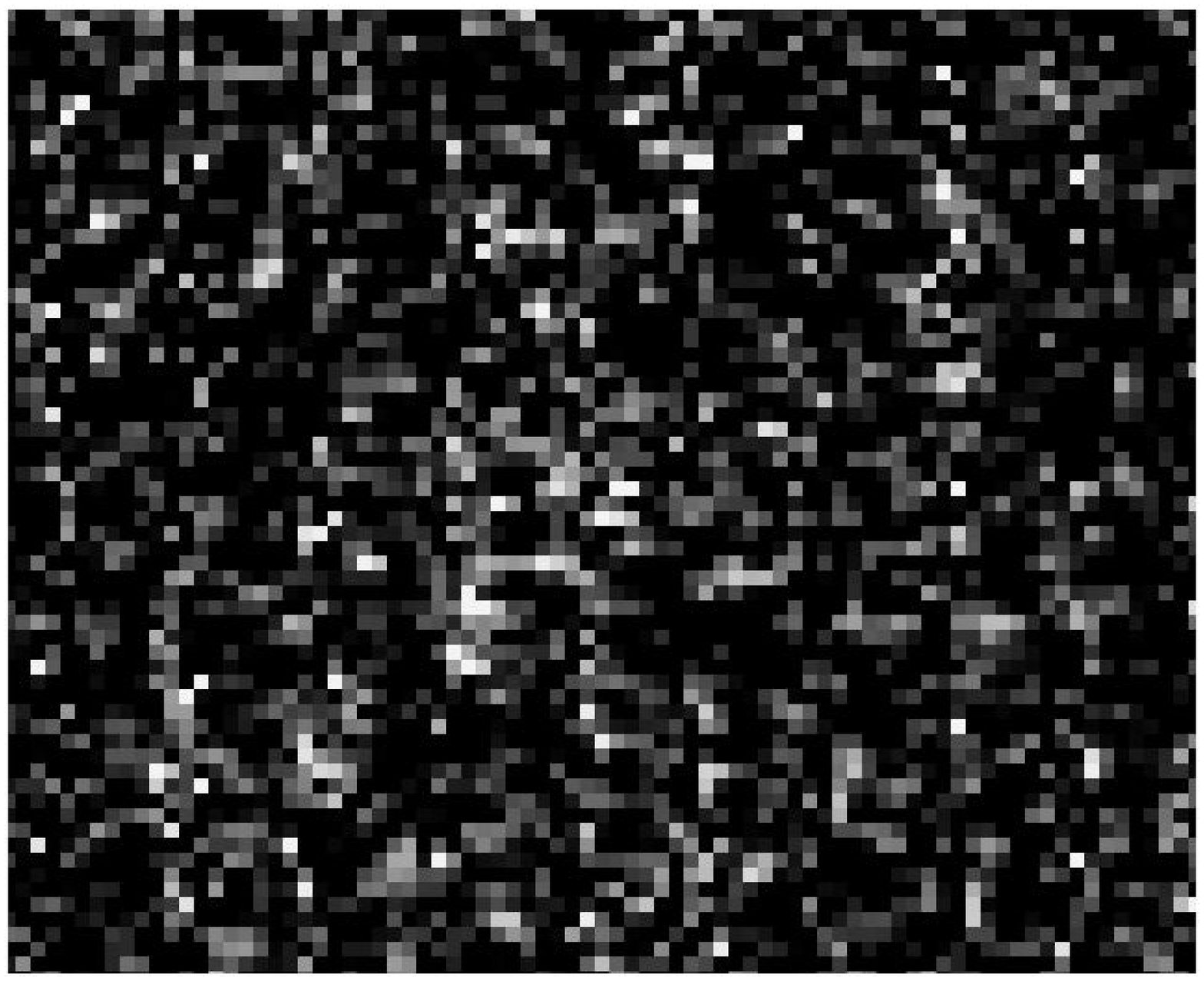} 
\caption{ACS HRC F330W image (left panels), GALFIT model (center panels) and residuals (right panels) of our sample. Arrows indicate the direction of the radio source. The dashed lines in {\object 1443+77} sketch the arc of emission mentioned in Section \ref{subsec:acsnotes}. \label{galfigs}}
\end{figure*}

The 2-D fitting code GALFIT \citep{Peng02} was used to parameterize the UV emission. For each image we tested different combinations of point source and Sersic profiles, allowing the sky level, position and magnitudes of all components, as well the index and effective radii of the Sersic components, to vary. The final model was chosen according to the lowest $\chi^2$ and best residuals (with the lowest number of components). The results are summarized in Table \ref{ACSPosi}. Figure \ref{galfigs} shows the UV image, GALFIT model and residuals for the nine objects. These will be discussed in detail in Section \ref{uvmorph}.

\begin{figure*}[h]
\centering
\includegraphics[width=0.5\columnwidth]{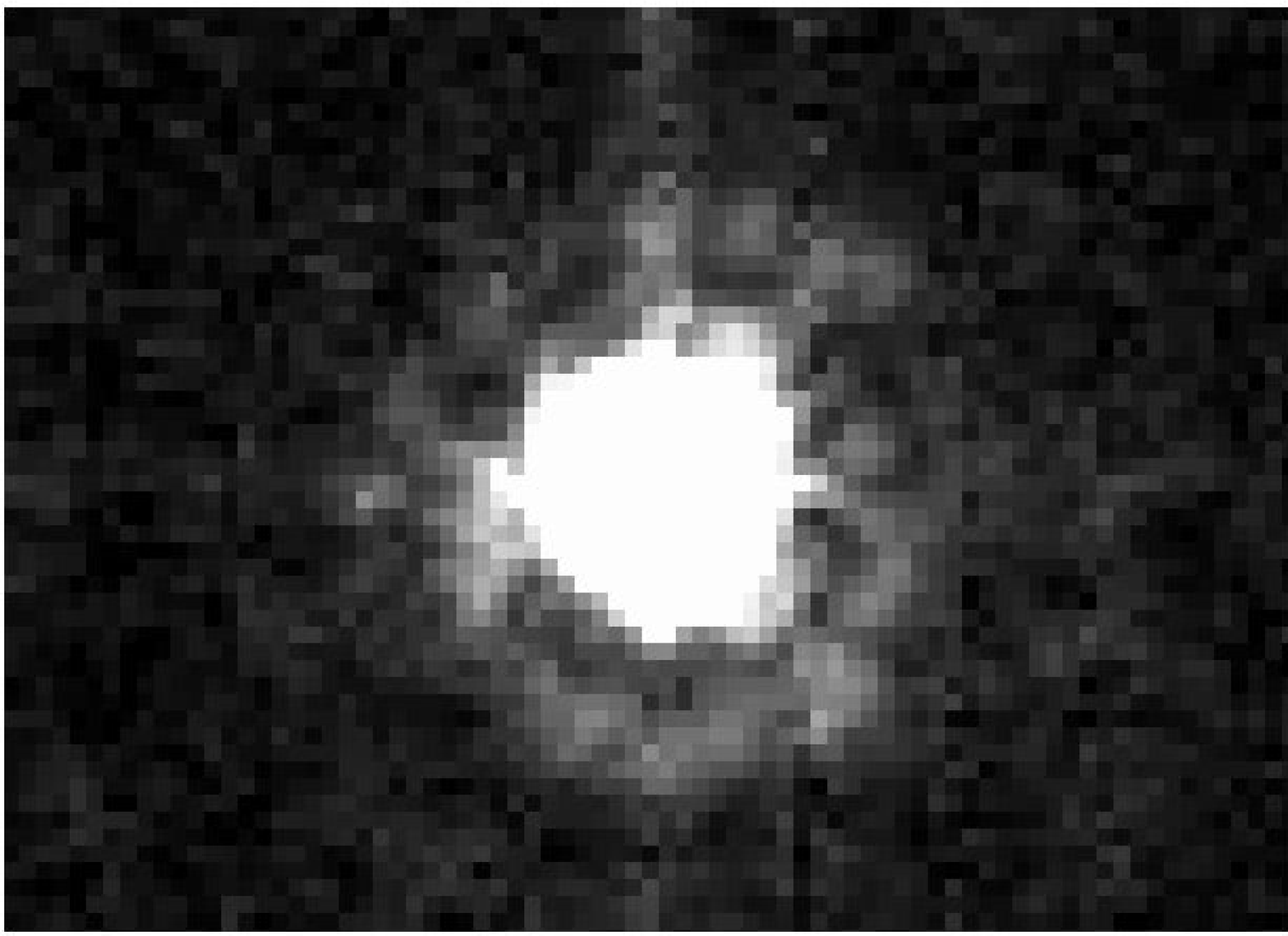} \hfil \includegraphics[width=0.5\columnwidth]{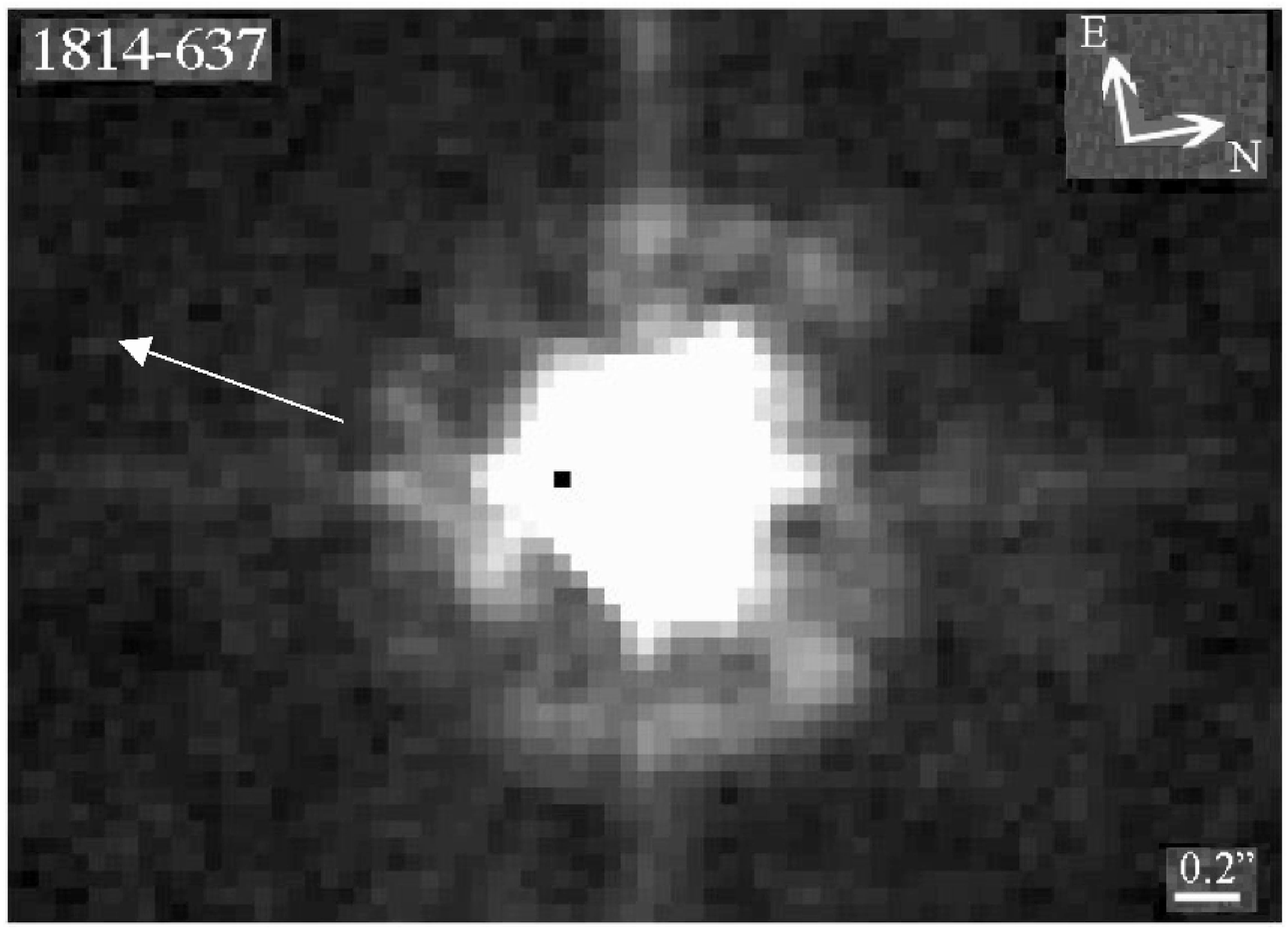} \hfil \includegraphics[width=0.5\columnwidth]{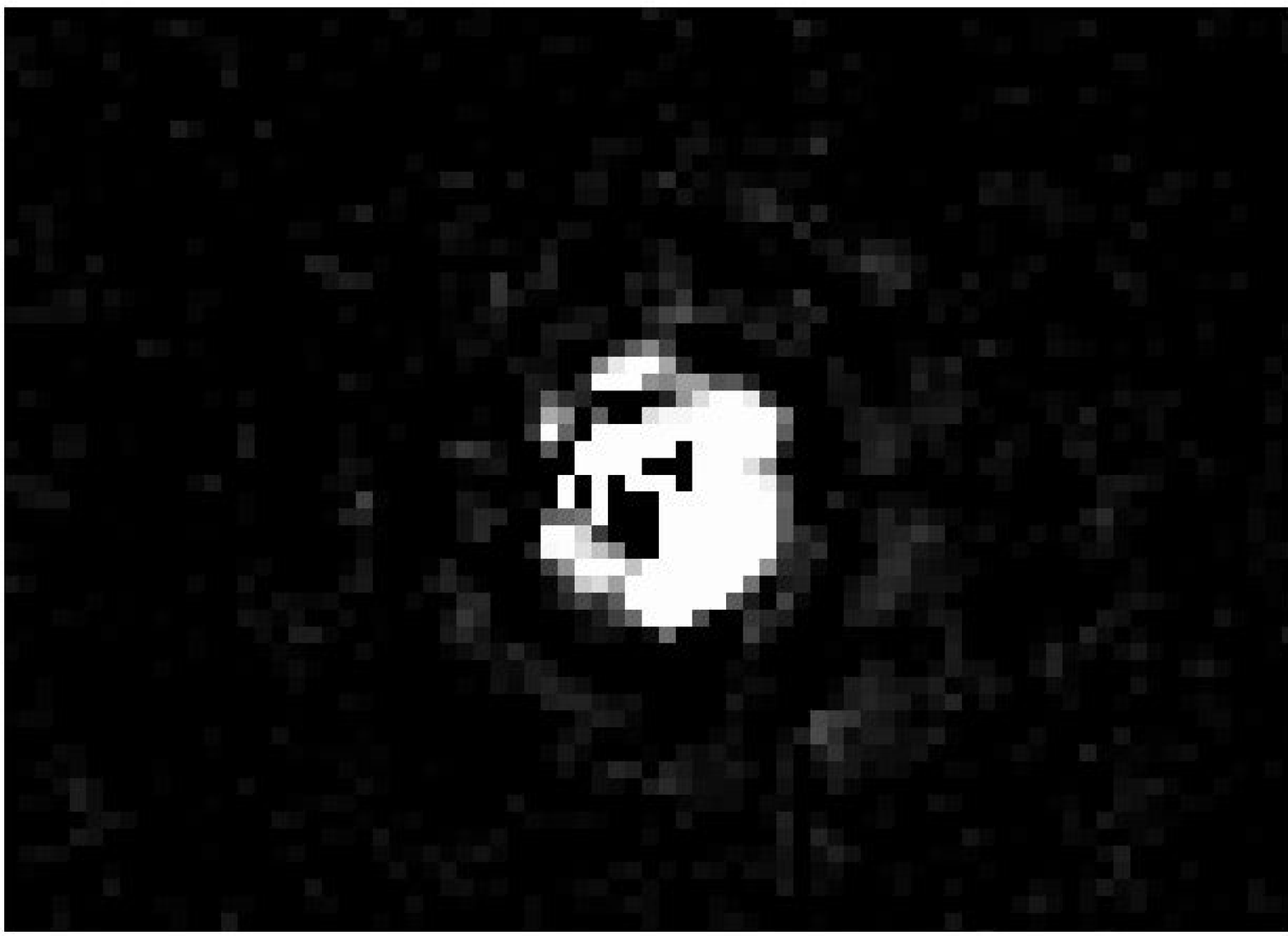} 

\includegraphics[width=0.5\columnwidth]{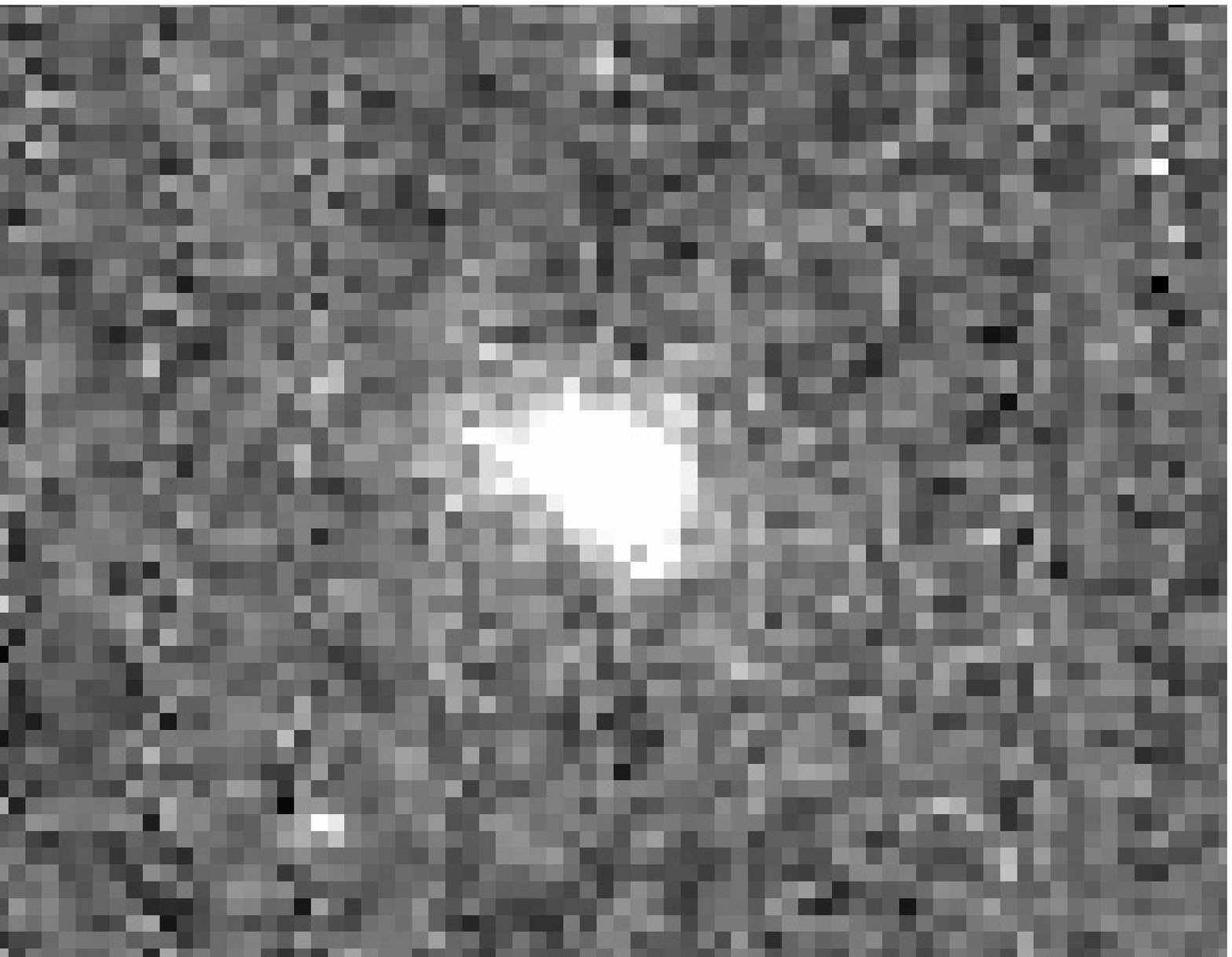} \hfil \includegraphics[width=0.5\columnwidth]{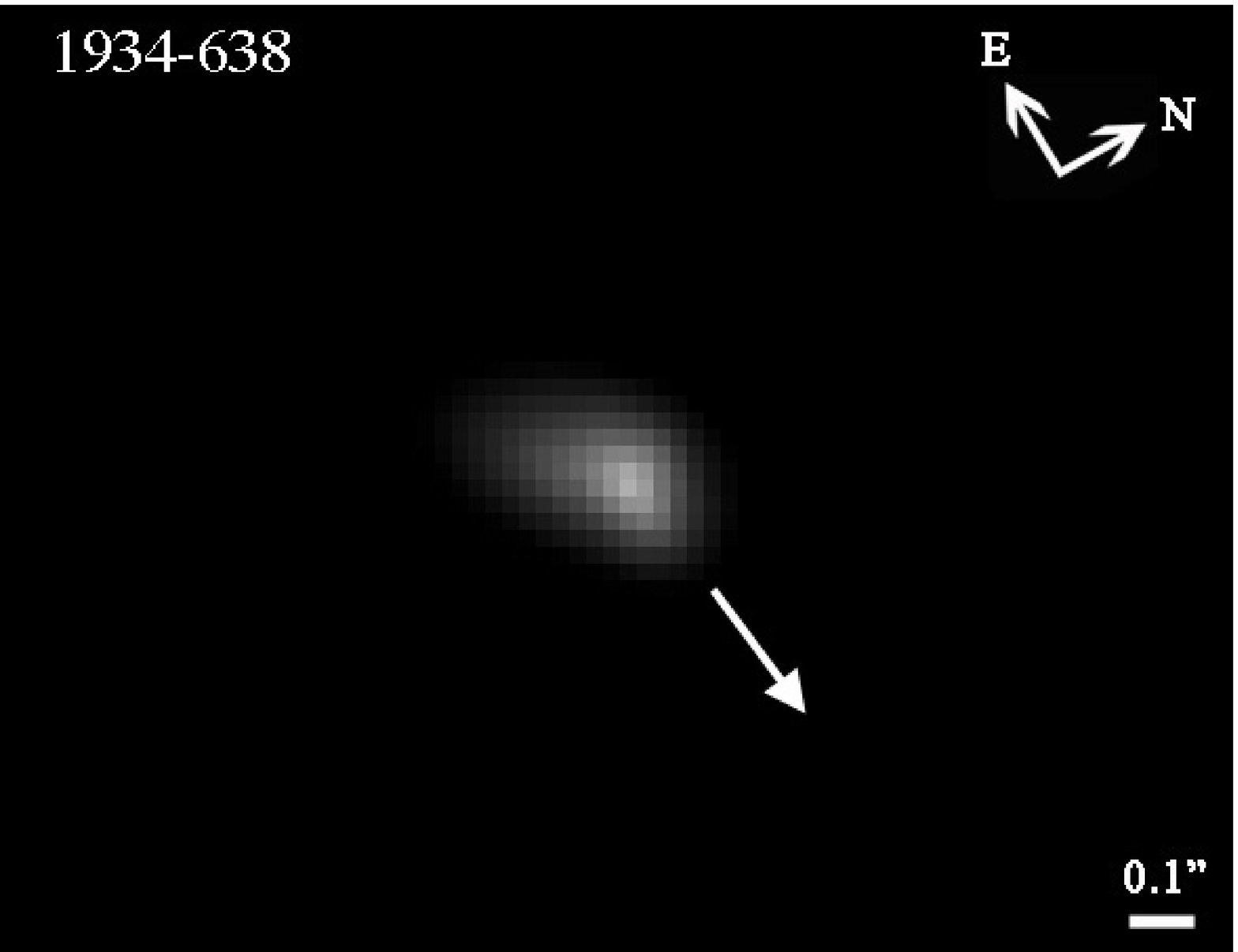} \hfil \includegraphics[width=0.5\columnwidth]{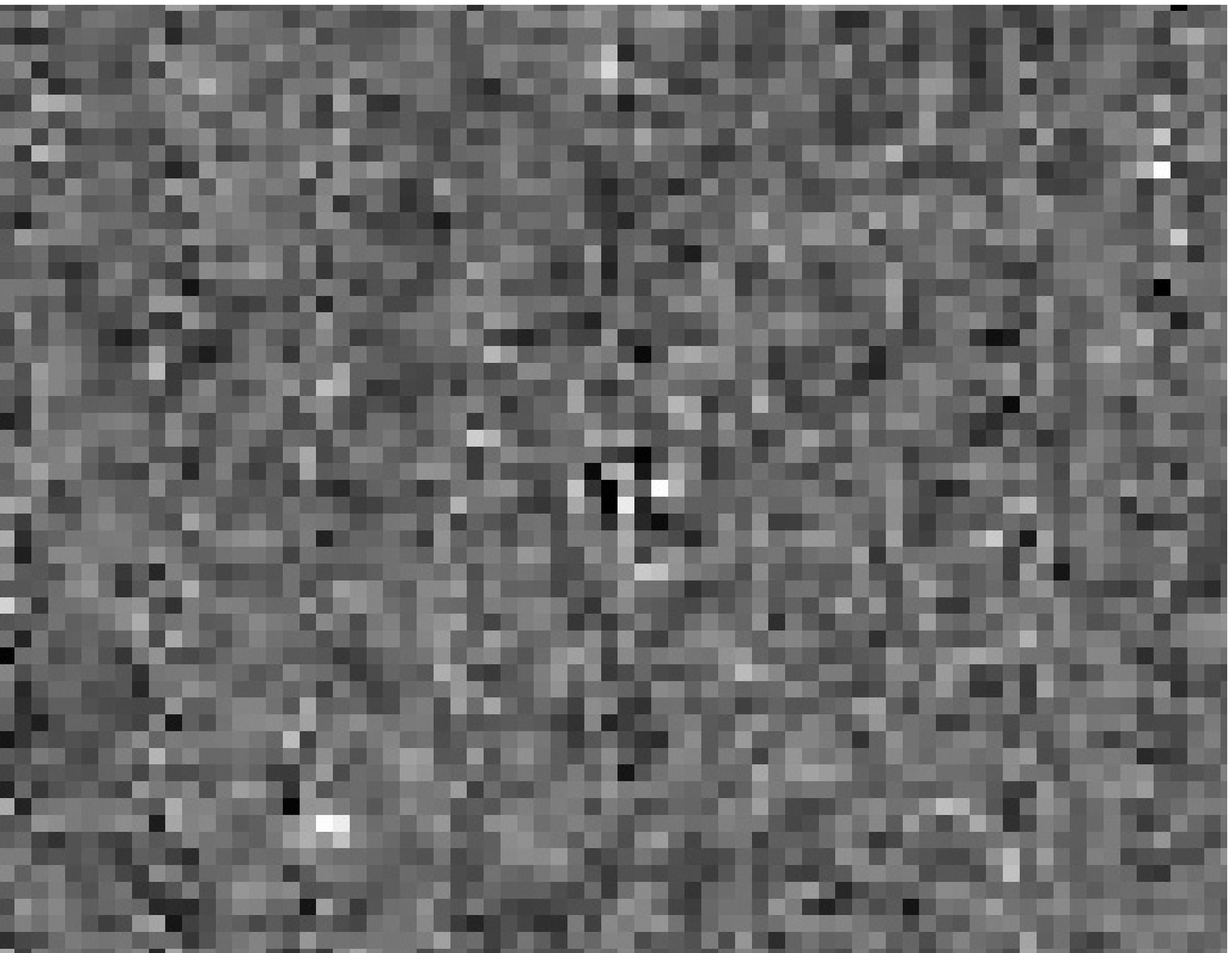} 

\includegraphics[width=0.5\columnwidth]{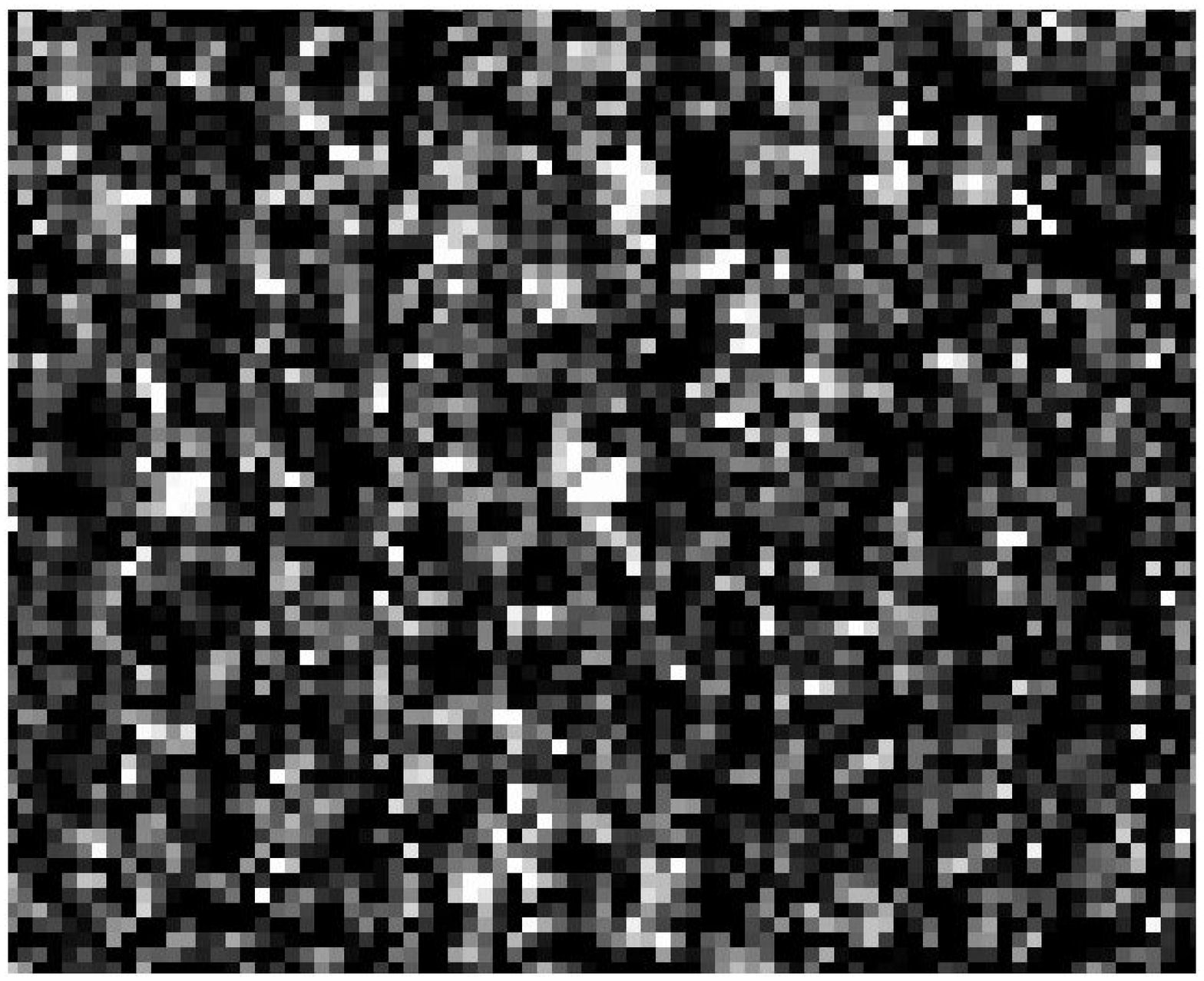} \hfil \includegraphics[width=0.5\columnwidth]{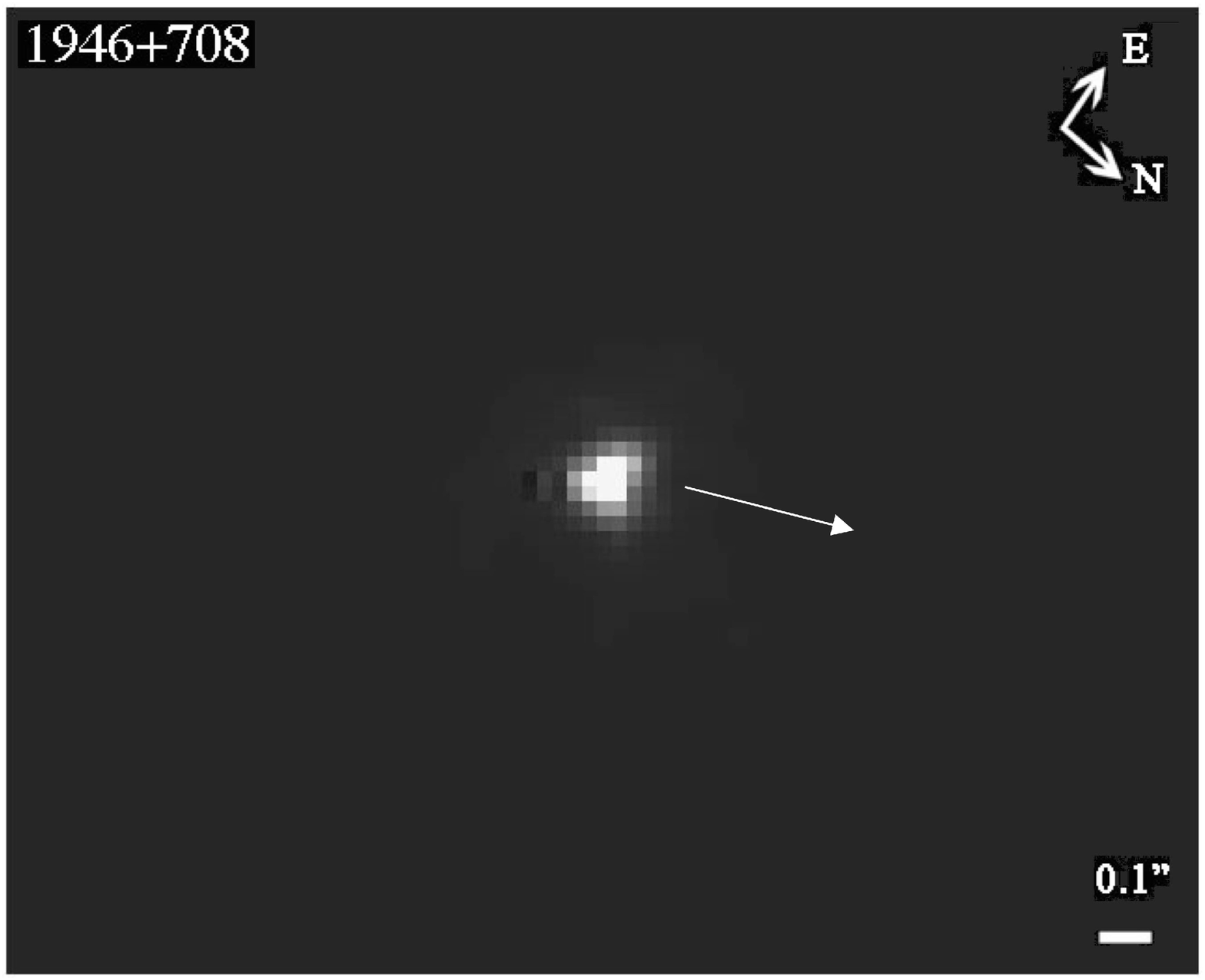} \hfil \includegraphics[width=0.5 \columnwidth]{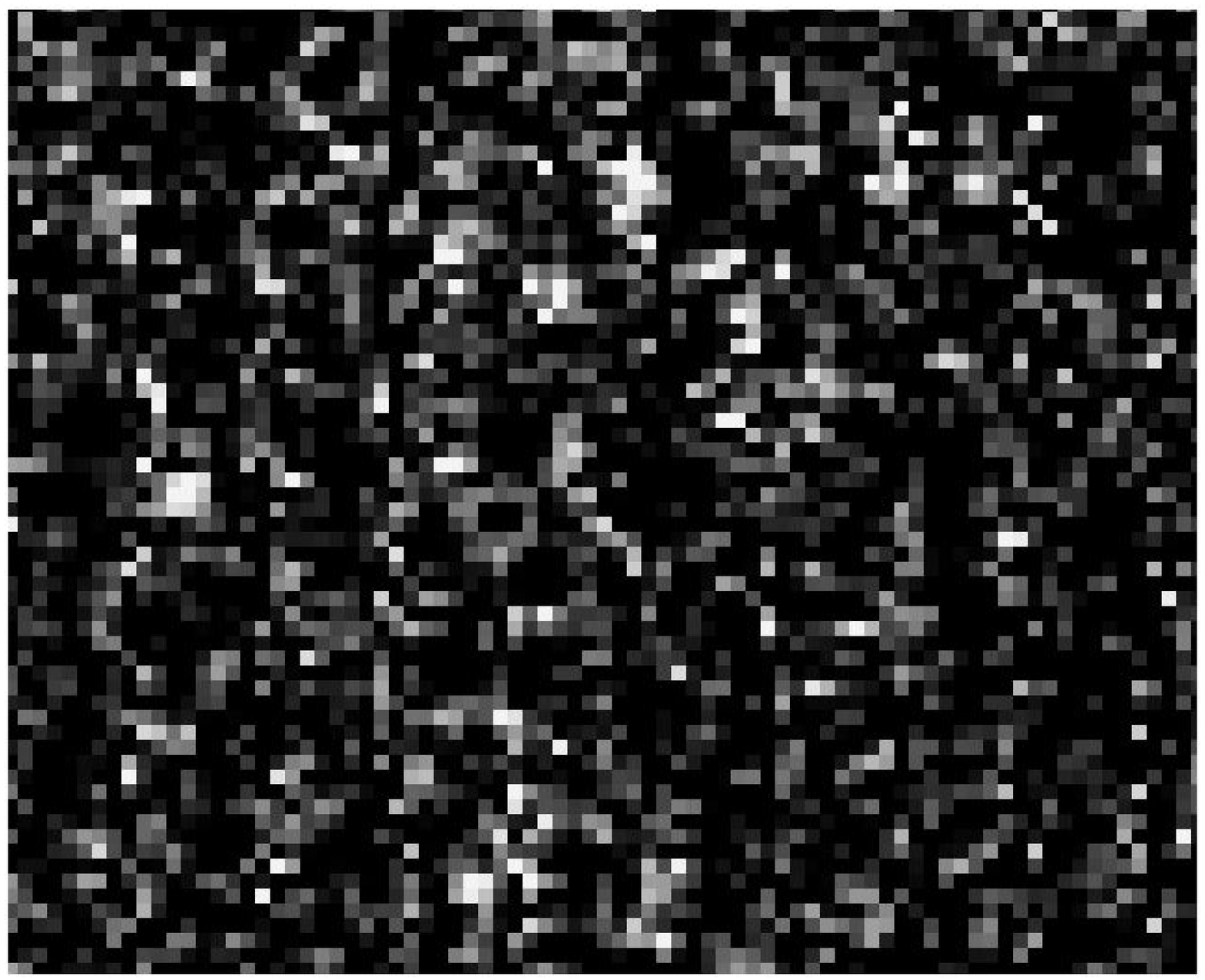} 

\includegraphics[width=0.5\columnwidth]{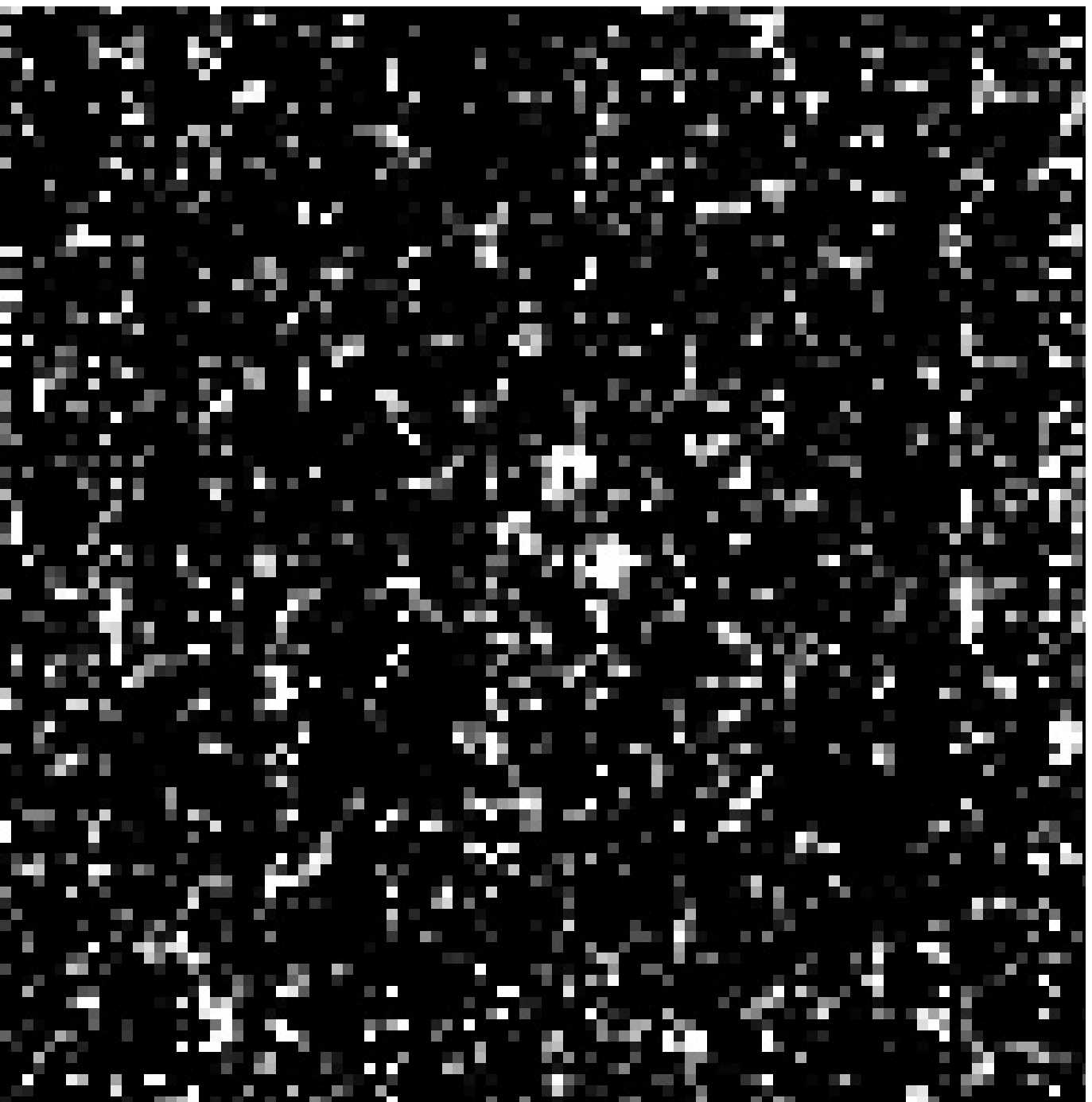} \hfil \includegraphics[width=0.5\columnwidth]{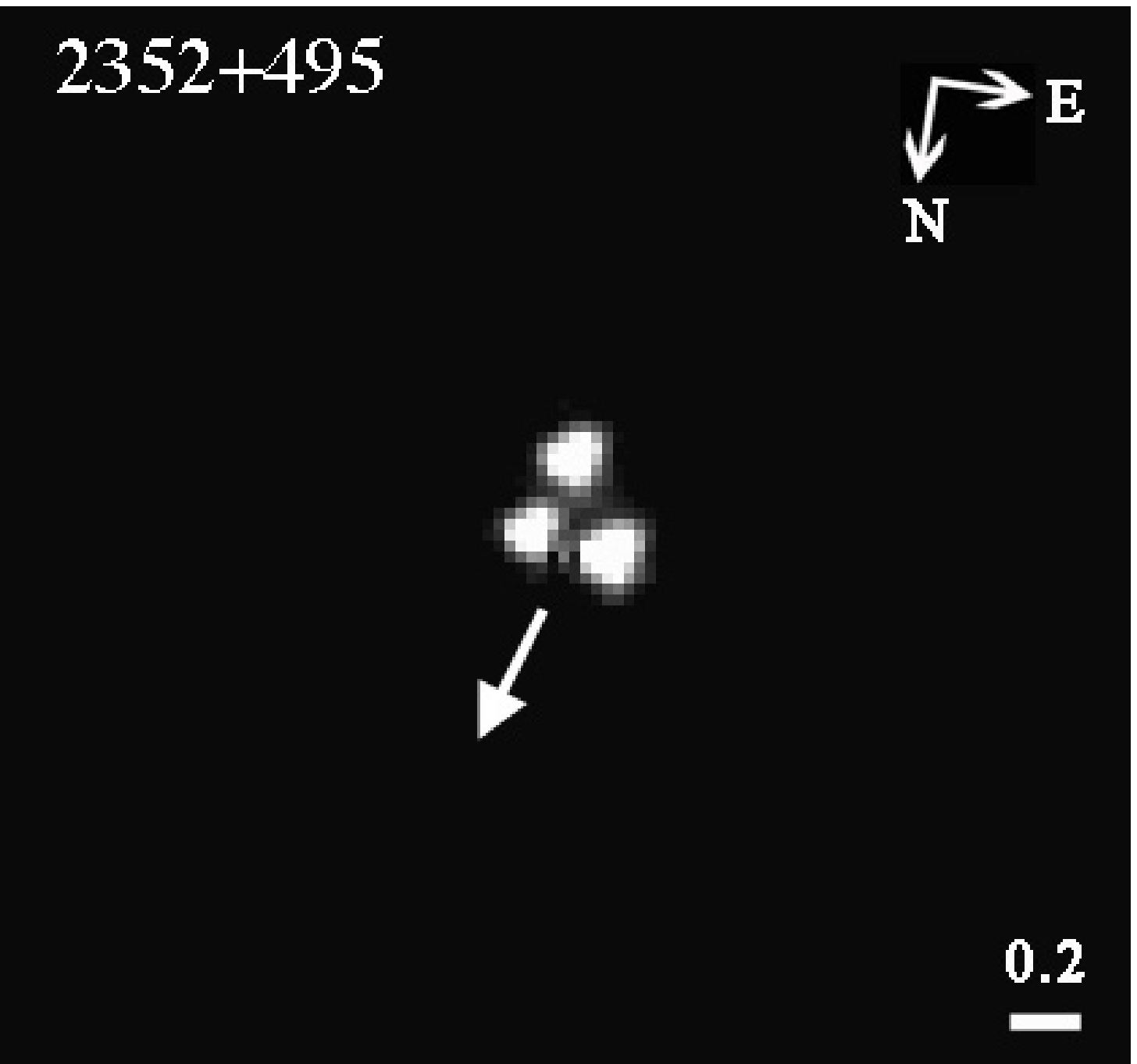} \hfil \includegraphics[width=0.5\columnwidth]{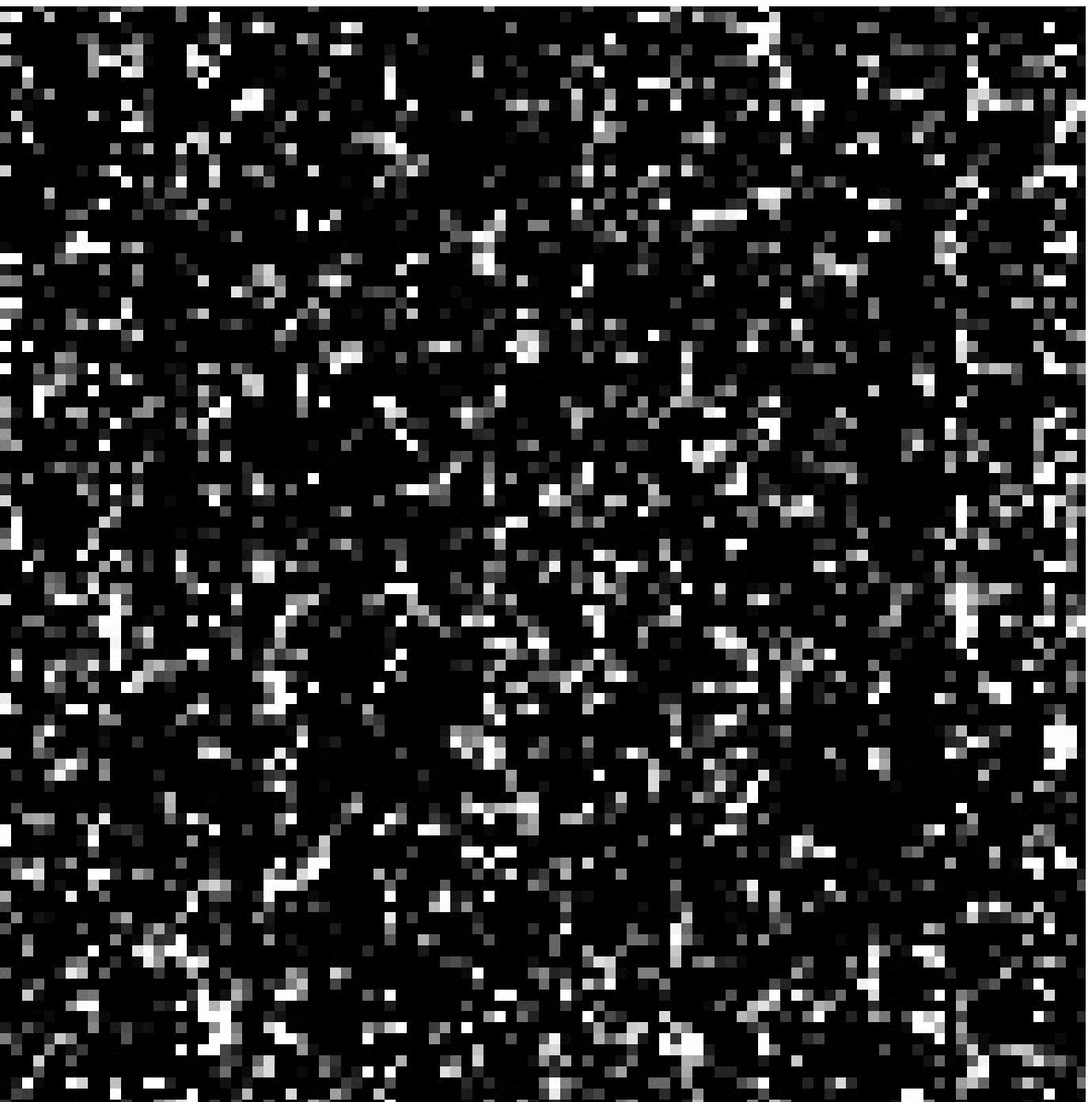} \\\

\begin{flushleft}
{\bf Fig. \ref{galfigs}.} Continued.
\end{flushleft}
\end{figure*}

A Sersic profile \citep[][ for a review]{Sersic63, Sersic68, Graham05}, R$^{1/n}$ is described by:

\begin{equation}
I(R) = I_e \, exp\left[ -b_n \left( \left( \frac{R}{R_e}\right)^{1/n} - 1 \right) \right]
\end{equation}
where $I_e$ is the intensity at the effective radius, $R_e$ (distance that encloses half of the total emission), and $b_n$ is a constant coupled to the value of $n$. Special cases of the Sersic profile are those where $n=4$ (de Vaucouleurs profile), $n=1$ (exponential profile) and $n=0.5$ (Gaussian profile).

The TinyTim (Space Telescope Science Institute program to generate simulated HST
point spread functions) models for the HRC PSF are not sufficient for our purposes.
Instead,  we used the {\it calibration plan} observations of Cycles 12 and 13 (programs 10054 and 10374). These programs contain observations of the spectrophotometric standard stars \object{GD71}, \object{G191B2B}, \object{GD153} and \object{HZ~14}. To model our PSF, we compared the PSF created by each of these stars in each Cycle, as well as combinations of these. In general, the differences between each PSF are subtle and do not produce major differences in the models of the sources. However, the average PSF of the spectrophotometric stars of Cycle 13 produced better point source results (lower $\chi^2$ and residuals) so we chose to use it in our models. This PSF has a FWHM of $\sim$1.9 pixels, implying a resolution of $\sim 0.05\arcsec$. However, the residuals for \object{1814--637} suggest
that an even better PSF model is required (see Figure \ref{galfigs}).


GALFIT yields the coordinates of the modeled components in pixels. The conversion to RA and Dec was performed using the astrometry information stored in the header of each HST/ACS -pipeline reduced- image (model errors are listed in Table \ref{ACSPosi}). The radio positions of the sources are from the literature. However, the registration between HST and radio frames can be off by as much as 1$\arcsec$. 

Unless otherwise noticed, all presented fluxes, luminosities and magnitudes are for the F330W filter passband, corrected for galactic extinction and k-correction. The magnitudes are in the STMAG system. Galactic extinction was corrected using the Galactic de-reddening curve in \citet{Cardelli89} and the measured Galactic extinction values of \citet{Schlegel98}. K-correction (typically $\sim$ 0.1, 0.2 magnitudes) and conversion of the \cite{Allen02} filters were done following the PEGASE \citep{Fioc97} templates and using the IRAF package SYNPHOT. SYNPHOT was also used for measuring F330W magnitudes of the PEGASE and GALAXEV \citep{Bruzual03} stellar population models.
We use H$_0$=71, $\Omega_M = 0.27 ,  \Omega_\Lambda = 0.73$ \citep{Spergel03} throughout the paper.

\section{Nuclear and emission line contamination}

Here we discuss several possible contributions to the near-UV
light in these radio galaxies.
Using ground-based, optical spectroscopy, \citet{Tadhunter02} studied the
nature of the UV excess in GPS, CSS and FR II sources at redshifts
0.15 $<$ z $<$ 0.7. They found that the UV continuum in these sources
has contributions from (1) nebular continuum, (2) direct AGN light,
(3) scattered AGN light, and a (4) starburst component. In addition to these continuum components
we also consider contamination from emission line gas.

The observed objects  are Narrow Line Radio Galaxies (NLRG) so we expect no contamination from {\it direct\/} light from the AGN. The main contribution from emission line gas to our observations would come from \ion{Mg}{ii} (see Table \ref{zlambda}). It is usually only found in the nuclear Broad Line Region (BLR) of AGN hosts so we do not expect direct
contamination from emission line gas.


To make a rough estimation of the extent of the possible contamination by direct light from the AGN, we use the STIS spectrum of \object{3C~277.1} \citep{Labiano05}, a CSS QSO that could represent our worst case scenario: when we are looking into the nucleus. In \object{3C~277.1}, all signs from nuclear contribution (broad lines and AGN continuum) and \ion{Mg}{ii} emission disappear at ~0.8 kpc from the nucleus, roughly 1.5 times the FWHM of the PSF in the STIS observations. On average, 0.8 kpc correspond to $\sim0.2\arcsec$ in our ACS observations and we could expect the nuclear traces to disappear closer to the center in our ACS galaxies.  The \ion{C}{ii}] line is usually fainter than \ion{Mg}{ii} \citep{Peterson97} in AGN (it is also not present in \object{3C~277.1}) so it is unlikely that it is affecting our observations.

There are UV polarization observations of only two objects, \object{1934--638} and
\object{1345+125}, so the presence of
scattered nuclear light in the sample cannot be completely ruled out.
\citet{Tadhunter02} find AGN scattered light in $\sim 37\%$ of their sample
(including the GPS \object{1934--638}), but in most cases it does not
seem to dominate the UV emission. As we note below in Section 5.1,
the UV light in the GPS sources tends not to be aligned with the radio source
suggesting that it is not due to scattered nuclear light (assuming the scattered photons
would escape along the radio axis). 
Observations of \object{1345+125} by  \citet{Hurt99} are consistent with the
existence of polarized UV light (though with large uncertainty).
However, \citet{Tadhunter05} found evidence of recent star formation and
strong jet cloud interactions in \object{1345+125}
\citep[see also][]{Holt03, Surace98}.

In general, the contribution from nebular continuum in radio galaxies varies
between 3 and 40$\%$ \citep{Tadhunter02}. However, 3 out of the 4 CSS
they studied are dominated by young stellar populations.
Furthermore, for up to 50$\%$ of their sources (including GPS, CSS and FR II),
the UV excess is dominated by young stellar populations. Similar results
have been found for FR I and FR II sources at a range of redshifts
\citep{Aretxaga01, Willis02, Willis04}.

To sum up, the most probable sources of UV light in our images are
 star formation, scattered nuclear light and nebular continuum. However, it is
likely that the contributions of the latter two are small.

\section{UV morphology}
\label{uvmorph}


\begin{table}[t]
\caption{Redshift and wavelengths.}
\label{zlambda}
\begin{minipage}{\columnwidth}
\centering
\resizebox{\textwidth}{!}{
\begin{tabular}{cccccc}
\hline
\hline
Source   & Catalogue & ID & Sample & Redshift &  Emission lines \\
\hline 
\object{1117+146} & \object{4C 14.41} & G & GPS & 0.362 & \ion{C}{ii}] 2326, \ion{Mg}{ii} 2800 \\

\object{1233+418} &  & G & CSS & 0.25 & \ion{Mg}{ii} 2800\\

\object{1345+125} & \object{4C 12.50} & G & GPS & 0.12174 &  \ion{Mg}{ii} 2800\\

\object{1443+77}   & \object{3C~303.1} & G & CSS & 0.267 &   \ion{C}{ii}] 2326, \ion{Mg}{ii} 2800\\

\object{1607+268} & \object{CTD~093} & G & GPS & 0.473 &  \ion{C}{ii}] 2326 \\

\object{1814--637}& & G & CSS & 0.063 &  [\ion{Ne}{V}] 3426 \\

\object{1934--638} & & G & GPS & 0.183 & 2 \ion{Mg}{ii} 2800\\

\object{1946+708} & & G & GPS & 0.10083 &  \ion{Mg}{ii} 2800\\

\object{2352+495} & \object{DA~611} & G & GPS & 0.23790 &  \ion{Mg}{ii} 2800 \\

\hline
\end{tabular}
}
\end{minipage}
\\ \\
B1950 IAU and catalogue names, identification, GPS/CSS classification, redshift and possible emission lines affecting our measurements.


\end{table}


\begin{table*}[t]
\caption{Galfit components.} 
\label{ACSPosi}
\begin{minipage}{\columnwidth}
\centering
\resizebox{2\textwidth}{!}{
\begin{tabular}{clllcccccc}
\hline
\hline
Source   &  Component & RA (J2000) & Dec (J2000) & STMAG & R$_e$ (mas) & R$_e$ (pc) & Index & Ratio & PA \\
\hline 
\object{1117+146} &\cite{Fey04} &11:20:27.807 $\pm$0.001 & +14:20:54.99 $\pm$0.02 \\
		 & Point source & 11:20:27.7653 $\pm$0.0001 & 14:20:54.33 $\pm$0.06 & 24.06$\pm$0.06 \\

\object{1233+418} & \cite{Becker95} &12:35:35.71 $\pm$0.03 & +41:37:07.40 $\pm$0.32\\
		 & Point source (?) & 12:35:35.6664 $\pm$0.0004 & +41:37:08.18 $\pm$0.17 & 24.93$\pm$0.11 \\

\object{1345+125} &  \cite{Ma98} & 13:47:33.3616 & +12:17:24.240 & \\
		 & Point source & 13:47:33.3981 $\pm$0.0001&+12:17:23.36 $\pm$0.04& 23.77$\pm$0.05 \\
		 & Point Source &13:47:33.3947 $\pm$0.0002&+12:17:23.46 $\pm$0.01& 21.49$\pm$0.01 \\ 
		 & Sersic profile &13:47:33.3950 $\pm$0.0001&+12:17:23.41 $\pm$0.06& 21.20$\pm$0.04  & 108 $\pm$7 & 234$\pm$15&1.62 $\pm$0.13 & 0.76$\pm$0.03& --39\\
		 & Sersic profile &13:47:33.5272 $\pm$0.0003&+12:17:23.24 $\pm$0.21& 21.84$\pm$0.12  & 299 $\pm$58 & 647$\pm$126& 2.29$\pm$0.38& 0.54$\pm$0.05& 19\\

\object{1443+77}   & \cite{Rengelink97} &14:43:14.9$\pm$1.1 & +77:07:28.6$\pm$3.8 \\
		 & Point source (?) & 14:43:14.666 $\pm$0.001& +77:07:27.546 $\pm$0.17& 25.21 $\pm$0.10 \\
		 & Point source & 14:43:14.5832  $\pm$0.0004& +77:07 27.702 $\pm$0.06& 24.06 $\pm$0.06 \\
		 & Sersic profile & 14:43:14.656 $\pm$0.002& +77:07:27.430 $\pm$0.10& 20.36 $\pm$0.16 & 1100$\pm$200&4500$\pm$800 & 2.66$\pm$0.33 & 0.45$\pm$0.02 & 83\\ 

\object{1607+268} & \cite{Beasley02} & 16:09:13.3208 & +26:41:29.036 \\
		 & Point source & 16:09:13.2497 $\pm$0.0002& +26:41:29.514 $\pm$0.14& 24.38 $\pm$0.09\\
		 & Point Source & 16:09:13.2472 $\pm$0.0002& +26:41:29.515 $\pm$0.14& 23.92 $\pm$0.05 \\ 

\object{1814--637}& \cite{Ma98} & 18:19:35.003 $\pm$0.003 & --63:45:48.194 $\pm$0.01   \\
		 & Point source & 18:19:35.3179 $\pm$0.0001& --63:45:47.117 $\pm$0.01& 16.38$\pm$0.10 \\

\object{1934--638} & \cite{Ma98} & 19:39:25.027 $\pm$0.001 & --63:42:45.626 \\
		 & Point source & 19:39 25.0947 $\pm$0.0009& --63:42:44.978 $\pm$0.13 & 23.81$\pm$0.73 \\
		 & Sersic profile & 19:39:25.0935 $\pm$0.0001&--63:42:44.974 $\pm$0.02 & 21.75$\pm$0.12 & 39$\pm$5 &119$\pm$15&1.02$\pm$0.33 & 0.13$\pm$0.04 & 38\\
		 & Sersic profile & 19:39:25.1032  $\pm$0.0007& -63 42:45.037 $\pm$0.06 & 21.08$\pm$0.07 & 354$\pm$45 & 1078$\pm$137 & 2.51$\pm$0.26 & 0.47$\pm$0.02 & 68\\

\object{1946+708} & \cite{Beasley02} & 19:45:53.5200 $\pm$0.002 &+70:55:48.732  \\
		 & Point source (?) & 19:45:53.3016 $\pm$0.0006& +70:55:48.476 $\pm$0.12 & 24.53$\pm$0.08 \\

\object{2352+495} &\cite{Ma98} & 23:55:09.458 $\pm$0.001 & +49:50:08.340 $\pm$0.01 \\
		 & Point source & 23:55:09.4704 $\pm$0.0004& +49:50:07.326 $\pm$0.16& 25.13$\pm$0.16 \\
		 & Point source (?) & 23:55:09.4634 $\pm$0.0006& +49:50:07.507 $\pm$0.25& 25.53$\pm$0.19 \\
		 & Point source & 23:55:09.4828 $\pm$0.0003& +49:50:07.531 $\pm$0.12& 24.79$\pm$0.10 \\
\hline
\end{tabular}
}
\end{minipage}
\\ \\
The first line for each source corresponds to the most recent radio position, the rest of the lines correspond to the UV components of the GALFIT models. Errors not listed are smaller than 0.1 milliarcsec. The coordinates in the GALFIT models are from the HST coordinate system and correspond to the brightest pixel. The magnitudes are observed, not corrected from galactic extinction. R$_e$ is the effective radius in milliarcsec. Last three columns give the Sersic indices axial ratios and position angles. The ``?'' marks marginal detections.
\end{table*}

We have modeled the UV emitting regions with GALFIT and found that  
\object{1117+146} and  \object{1814--637}.  show one point source component. 
There is a marginal detection of a point source in \object{1233+418} and \object{1946+708}.
The hosts of \object{1345+125}, \object{1443+77} and \object{1934--638} show a combination of at least one Sersic\footnote{The Sersic profiles are used to parameterize the data. It does not necessarily imply that these UV components are galaxies.} component (with different indices) and one or several point sources (see Table \ref{ACSPosi} and Figure \ref{galfigs}). \object{1607+268} and \object{2352+495} show a combination of two and three point sources respectively. In the near-IR, \citet{Vries00} find Sersic indices 2 $\lesssim$ n $\lesssim$ 5 and effective radii 2 kpc $\lesssim$ R$_e \lesssim$ 4 kpc for GPS and CSS sources. Most of our data show point sources and/or Sersic profiles with indices n $\lesssim2$ and radii R$_e\lesssim400$ parsec. The presence of these small clumps of near-UV emission is consistent with star forming regions in the host. Before addressing the nature of the UV emission, we describe the properties of this emission in relation to other properties, 
 for the individual sample sources. We note that optical observations of the higher redshift 
Parkes half-Jansky sample suggest that the rest frame UV may contain a contribution from a young 
stellar population (de Vries et al. 2007).

For our observations, the 3$\sigma$ detection limit, for a point source (FWHM $\sim$3 pixels) is 25.7 (no Galactic extinction applied), calculated from the noise in the images. The ACS Imaging Exposure Time Calculator gives the following  3$\sigma$ detection limits for our observations: Johnson U = 21/arcsec$^2$ for an extended uniform source (2x2 pixel extraction region) and 25.2 for a point source\footnote{Sources with a B star spectrum and fluxes $3.0 \times 10^{-17} $ ergs cm$^2$/s/\AA/arcsec$^2$ (extended source) and 6.0$\times 10^{17}$ ergs cm$^2$/s/\AA ~(point source), at 3000\AA.}. For comparison purposes with this limit and published data, the magnitudes listed in this section and Table \ref{ACSPosi}  have not been corrected for Galactic extinction.

\subsection{Notes on individual sources}
\label{subsec:acsnotes}

{\bf \object{1117+146}}: Identified as a GPS by \citet{O'Dea91}. The counterpart of the radio source corresponds to a m$_R$=20.1 galaxy \citep{Vries95} at z = 0.362 \citep{Vries98}. Radio observations \citep[e.g.,][]{Fey97} show a $\sim$100 mas double radio source. Our image shows an unresolved 24.06 magnitude source.

{\bf \object{1233+418}}: A CSS galaxy with a photometric R-band redshift z$_\mathrm{R}$ = 0.25 \citep{Fanti04}. We have a marginal detection of a point source with magnitude 24.93. Our image (see Figure \ref{galfigs}, first panel) suggests faint extended emission towards Northeast but the errors in the GALFIT model for this component are too large to be sure of its existence. Checking the individual pre-drizzle images, it is not clear if these components are real or artifacts.

{\bf \object{1345+125}}: \object{4C~+12.50}. A long known peaked spectrum radio source \citep[e.g.,][]{Veron71}. The counterpart of this GPS source is a
m$_R$=15.5 galaxy \citep[e.g.,][]{Stanghellini93} at z = 0.12174 \citep{Holt03}, in a cluster of fainter galaxies \citep{Stanghellini93}. IR images of this well known ULIRG show an extremely reddened source with two nuclei separated $\sim 1.8\arcsec$ ($\sim$4 kpc) embedded in a common envelope and aligned roughly East-West \citep[e.g.,][]{Scoville00, Surace00} suggesting an ongoing merger, which may have triggered the AGN \citep[e.g.,][]{Heckman86, Xiang02}. The western nucleus is the brightest and shows a Seyfert 2 spectrum \citep{Gilmore86}. \citet{Veilleux97} suggest that the source may have a hidden quasar, also supported by UV polarized continuum emission \citep[p=16.4$\% \pm2.6\%$, the polarization vector approximately perpendicular to the radio axis][]{Hurt99}. VLBI imaging shows a complex, distorted $\sim$100 mas ($\sim$0.2 kpc) source \citep[e.g.,][]{Lister03}, roughly oriented North-South (PA$\sim-20\deg$). Older observations of \object{1345+125} related the radio s
 ource to the East nucleus but improved astrometry showed that it is related to the western one \citep[e.g.][]{Stanghellini97, Axon00, Fanti00}. \citet{Evans99} study the molecular gas in \object{1345+125} and suggest that the molecular gas is fueling the AGN. Our UV image detects both nuclei separated by $\sim 1.8\arcsec$. The East component was modeled with a Sersic profile with index $\sim$2.3 and an effective radius $\sim 0.3\arcsec$ and magnitude 21.84. The West nucleus shows a more complex structure: an extended  component with magnitude 21.20 and Sersic index 1.62 and a point source with magnitude 21.49. This more complex structure could be due to interaction with the radio source, the Eastern component or a combination of both. Optical emission line images \citep{Axon00, Batcheldor06} show an arc of emission $\sim 1\arcsec$ North of the Western nucleus and fainter emission at $2\arcsec$. We do not detect UV emission associated with these emission line features. They a
 lso detect a faint tail of emission stretching from the West nucleus towards the west, present in our image and not modeled by GALFIT (see Figure \ref{galfigs}). \citet{Surace98} detect {\it compact blue knots} in \object{1345+125} around the source which they attribute to star forming regions. We detect the southern knots $\sim 2\arcsec$ south of the nuclei  but not the northern one (see Figure \ref{galfigs}). 





{\bf \object{1443+77}}: \object{3C~303.1}. The optical counterpart of this
CSS corresponds to a galaxy of m$_v\sim$20 \citep[e.g.,][]{Sanghera95}, at
z = 0.267  \citep{Kristian78}. The MERLIN map \citep{Sanghera95} shows a
$\sim 1.8\arcsec$ long ($\sim$7 kpc) double radio source aligned NW-SE
(PA $\sim 47\deg$) aligned with the inner emission line gas
\citep[e.g.][]{Vries99}. XMM-Newton observations detect the ISM of
the host galaxy as well as a second component which could be either
Synchrotron Self Compton from the Southern radio lobe or hot gas shocked by
the expansion of the radio source \citep{O'Dea06}.
The center of \object{1443+77} shows a complex
structure in the optical, which could be due to an ongoing merger
\citep[e.g.,][]{Vries97, Axon00} and it also shows up in our UV image.
\citet{Axon00} and \citet{Vries99} find an arc of emission south of the nucleus
of the source and \citet{McCarthy95} detects circumnuclear [\ion{O}{iii}] as
far as 3$\arcsec$ from the center. Our image suggests the presence of an arc
of emission $\sim 0.9\arcsec$ ($\sim$4 kpc) from the nucleus. The integrated
magnitude of this arc is $\sim$23. A region of the sky with the same area
has a magnitude $\sim$25. The GALFIT model of the source consists of
two point sources of magnitudes 25.21 (marginal detection) and 24.06, and an extended
component with  Sersic index 2.66 and effective radius 1.1$\arcsec$ ($\sim$4.5 kpc). The
arc is not modeled by GALFIT so it is visible  in the residuals of Figure
\ref{galfigs}. \object{1443+77} is the best candidate in our sample to
be undergoing jet induced star formation (see below).

{\bf \object{1607+268}}: Also known as CTD93. VLBI maps \citep[e.g.][]{Dallacasa98} show a two-component, $\sim$60 mas, GPS source. The counterpart is a galaxy at z = 0.473 \citep{O'Dea91} with m$_\mathrm{r}$ = 20.4 \citep{Stanghellini93}. We observe two point sources of magnitudes 24.38 and 23.92. 

{\bf \object{1814--637}}: VLBI imaging shows a two component CSS galaxy \citep[e.g.,][]{Tzioumis02}. Although the radio source is small ($\sim 0.4\arcsec$, $\sim$0.3 kpc) its radio spectrum does not peak at $\sim$1 GHz. The optical identification corresponds to a m$_v$=18.0 galaxy at z = 0.063 \citep{Wall85}. We observe a bright, 16.38 magnitude point source. \citet{Danziger79} describe \object{1814--637} as a narrow line radio galaxy with a bright star-like nucleus. They report \ion{H}{$\alpha$} and \ion{H}{$\beta$} absorption lines at zero redshift and suggest that there is a foreground star superposed on the galaxy. The UV object is $\sim$2 arcsec to the Northeast of the radio source \citep{Ma98}. However, the registration between HST and radio frames can be off by as much as 1$\arcsec$, so it is probably the counterpart.

{\bf \object{1934--638}}: A long known GPS \citep{Bolton63}. VLBI maps \citep[e.g.,][]{Tzioumis98} show two components separated by $\sim 40$mas \citep[0.12 kpc at z = 0.183,][]{Tadhunter93}. $R$-band observations \citep{Jauncey86} show a system of two galaxies, separated by $\sim 3\arcsec$, consistent with our observations. The GALFIT model yields two extended components with  magnitudes 21.08 and 21.75  and a fainter point source with magnitude 23.81. \object{1934--638} shows significant polarization in the UV \citep[3.5$\%$, e.g.,][]{Tadhunter94,Morganti97}, with the position angle of the electric vector perpendicular to the radio axis. Scattered AGN light probably makes a significant contribution to the UV excess in this source \citep{Tadhunter02}. Optical polarization observations also suggest an anomalous environment \citep{Tadhunter94, Morgatnti97}.


{\bf \object{1946+708}}: The counterpart is identified with a z = 0.10083, m$_\mathrm{R}$=16.3 galaxy \citep[e.g.,][]{Snellen03b}. VLBI observations of this GPS source show an elongated, $\sim$40 mas ($\sim$ 0.07 kpc) NE-SW structure \citep{Taylor97}. Optical observations \citep{Perlman01} also suggest a NE-SW elongated source. Our near-UV image suggests a marginal detection of a point source with magnitude 24.53, present in two of the three individual pre-drizzle images. \citet{Perlman01} suggest that \object{1946+708} may be part of a group of galaxies and the closest would be $\sim1\arcmin$ far from \object{1946+708}. These objects fall outside our field and they have unknown redshifts.

{\bf \object{2352+495}}:  \citet{Snellen03b} identify the counterpart of this GPS source \citep[e.g.,][]{O'Dea98} with a z = 0.23790, m$_R$ = 18.2 galaxy. VLBI observations show a complex elongated $\sim 0.07\arcsec$ ($\sim$ 0.25 kpc) source oriented NW-SE \citep{Pollack03}. Our ACS observations show three point sources in a circle about 0.5$\arcsec$ wide, with magnitudes 24.79, 25.13 and 25.53 (slightly above the detection limit). The individual pre-drizzle show the two brightest sources while the third one is dubious.

\section{UV and radio properties}

Here we compare the UV and radio properties of our sample with those of  
the sample of 3CR FR I and FR II sources studied by \citet{Allen02}.  
The \citet{Allen02} data consist of STIS near-UV-MAMA snapshots (exposure time of 1440 seconds)
with filters F25SRF2  $\lambda_c = 2270 \AA$)
and F25CN182 ( $\lambda_c = 1820 \AA$). The radio data have been collected from the papers
listed in the individual notes above: \citet{O'Dea98}, \citet{Morganti93},
\citet{Martel99}, \citet{Flesch04} and the 3CRR on-line catalog
\citep[an update of the sample of][]{Laing83}. The data are summarized in
Table \ref{uvgraphs} and plotted in Figures \ref{UVRP} to  \ref{ALZ}. 


\begin{table*}[t]
\caption{UV and 5GHz radio properties.}
\label{uvgraphs}
\begin{minipage}{2\columnwidth}
\centering
\resizebox{\textwidth}{!}{
\begin{tabular}{ccccccccccc}
\hline
\hline
Name & Type & Size$_{5\mathrm{GHz}}$ &UV Size &	Log P$_{5\mathrm{GHz}}$   &Log(L$_{UV}$)	& L$_{UV}$ & PA 5GHz  & PA UV   &$\Delta$ PA \\ 
 & & (kpc) & (kpc)&	Log (Watts/Hz)    & Log L$_\odot$	&  10$^{40}$ erg/s & ($^o$) & ($^o$)&($^o$)\\ 
\hline 
\object{1233+418}  &CSS	&11.6	&1.2&25.56		&6.71$\pm$0.11 & 2.0$\pm$0.1&27 & -- & -- \\ 
\object{1443+77}  &CSS	&6.9	&8.1&25.86		&8.66$\pm$0.06 & 180$\pm$17&140 & 116, 136, 20$^a$& 4\\ 
\object{1814--637} &CSS	&0.50	&--&25.56		&8.83$\pm$0.03 & 265$\pm$20&156 & -- & --\\ 
\object{1117+146} &GPS	&0.40	&0.2&26.41		&7.45$\pm$0.02 & 11.0$\pm$0.4&120 &-- & -- \\ 
\object{1345+125} &GPS	&0.17	&0.45&25.98		&7.83$\pm$0.01 & 26.1$\pm$0.7&120, 26 &165& 45 \\ 
\object{1607+268} &GPS	&0.30	&0.2&26.86		&8.05$\pm$0.04 & 44$\pm$2&29 & 88 &59 \\ 
\object{1934--638} &GPS	&0.13	&0.5&26.65		&8.24$\pm$0.03 & 68$\pm$5&88 & 140 &52 \\ 
\object{1946+708} &GPS	&0.07	&0.2&25.14		&6.26$\pm$0.04 & 0.70$\pm$0.05&28 & -- &-- \\ 
\object{2352+495} &GPS	&0.86	&0.9&26.23	&7.39$\pm$0.03 & 9.6$\pm$0.7&153  &157, 27, 85$^b$ & 4\\ 
\object{3C~29} &	FR 1 &	133	&&24.99	&7.63$\pm$0.12&\\ 
\object{3C~35} &	FR 2 &	960	&&24.75	&7.42$\pm$0.45&\\ %
\object{3C~40} &	FR 2 &	20	&&24.12	&7.56$\pm$0.13&\\ 
\object{3C~66b} &	FR 1 &	300	&&24.45	&7.78$\pm$0.25&\\ 
\object{3C~192} &	FR 2 &	228	&&25.29	&8.16$\pm$0.17&\\ %
\object{3C~198} &	FR 2 &	512	&&24.79	&8.80$\pm$0.08&\\ 
\object{3C~227} &	FR 2 &	379	&&24.59	&9.27$\pm$0.08&\\ 
\object{3C~236} &	FR 2 &	4470	&&25.43	&8.41$\pm$0.35&\\ %
\object{3C~270} &	FR 1 &	55	&&23.74	&7.43$\pm$0.06&\\ 
\object{3C~285} &	FR 2 &	267	&&25.00	&8.40$\pm$0.06&\\ %
\object{3C~293} &	FR 1 &	225	&&24.91	&8.38$\pm$0.06&\\ %
\object{3C~296} &	FR 1 &	206	&&24.31	&7.78$\pm$0.08&\\ %
\object{3C~305} &	FR 1 &	11	&&24.59	&8.37$\pm$0.09&\\ 
\object{3C~310} &	FR 1 &	316	&&24.89	&7.23$\pm$0.14&\\ %
\object{3C~317} &	FR 1 &	18	&&24.37	&7.91$\pm$0.12 &\\ 
\object{3C~321} &	FR 2 &	540	&&25.33	&9.13$\pm$0.14&\\ %
\object{3C~326} &	FR 2 &	1990	&&24.83	&7.33$\pm$0.18&\\ %
\object{3C~338} &	FR 1 &	68	&&23.96	&7.56$\pm$0.04&\\ 
\object{3C~353} &	FR 2 &	94	&&24.57	&6.67$\pm$1.55&\\ 
\object{3C~382} &	FR 2 &	204	&&25.19	&10.35$\pm$0.22&\\ %
\object{3C~388} &	FR 2 &	84	&&25.48	&7.52$\pm$0.27&\\ %
\object{3C~390.3}&	FR 2 &	246	&&25.47	&9.44$\pm$0.22&\\ 
\object{3C~405} &	FR 2 &	138	&&27.15	&8.92$\pm$1.14&\\ 
\object{3C~449} &	FR 1 &	514	&&23.93	&7.98$\pm$0.51 &\\ %
\object{3C~465} &	FR 1 &	349	&&24.71	&7.94$\pm$0.22 &\\ %
\hline
\end{tabular}
}
\end{minipage}
\\ \\
UV luminosity for our GPS and CSS radio galaxies compared with large extended FR sources from \cite{Allen02}. Columns: (1) B1950 and 3C of the sources. (2) Classification. (3) Radio size. (4) UV size. (5) Radio power. (6) and (7) UV luminosity. (8) Position angle of the radio source. (9) Position angle of the UV source. (10) Difference in radio and UV position angles of the most aligned components. Longest linear sizes are in kpc. Radio and UV position angles in degrees (from North to East) of each component.The radio data are from \cite{O'Dea98}, \cite{Morganti93}, \citet{Martel99}, \citet{Flesch04} and the on-line 3CRR catalogue \citep{Laing83}, measured at 5GHz.\\  
{\small 
$^a$ The PA of the largest angular UV scale is 130$^{o}$.\\
$^b$ The UV morphology of  \object{2352+495} is a roughly equilateral triangle so no general PA for the complete source can be given (see Figure\ref{galfigs}). The PA's given are for the sides of the triangle formed
by the three point sources. The angles 157 and 85 are formed by the marginally detected component with the other two components.}
\end{table*}

We note that \citet{O'Dea97} found that the radio power of the bright GPS and CSS sample
is independent of size and comparable to that of the most powerful FR II sources.
However, the \cite{Allen02} FR sources are low z 3CR, and therefore lower radio luminosity  sources.

\subsection{GPS sources}

Most of the sources in our sample are very compact GPS sources, and therefore,  are probably too small
to strongly affect their environment on the scales resolved by these observations.
However, some conclusions can be drawn from their integrated UV and radio properties.

\begin{figure}[t]
\centering
\includegraphics[width=\columnwidth]{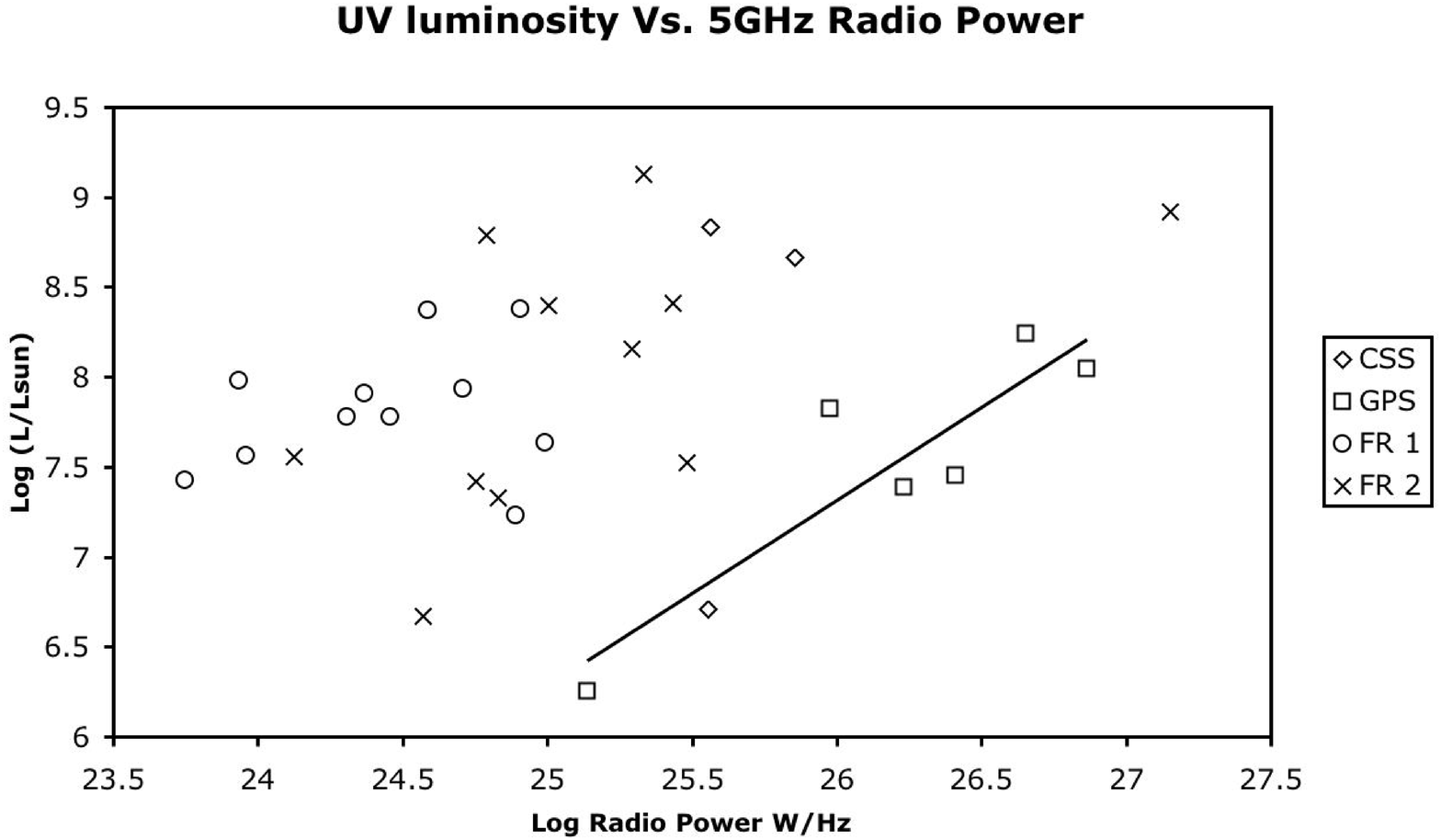}
\caption{UV luminosity and radio power of the combined sample of GPS/CSS and FR I, FR II from
\cite{Allen02}. The solid line shows the correlation between UV luminosity and radio power for GPS sources. \label{UVRP}}
\end{figure}

Figure \ref{UVRP} suggests a trend between UV luminosity and radio power of GPS sources \citep[consistent with the results of ][ for larger radio galaxies]{Raimann05}.  
This relationship could be produced by a larger reservoir of gas available in the more
luminous sources for fueling the radio source and supplying the starburst.

\begin{figure}[t]
\centering
\includegraphics[width=\columnwidth]{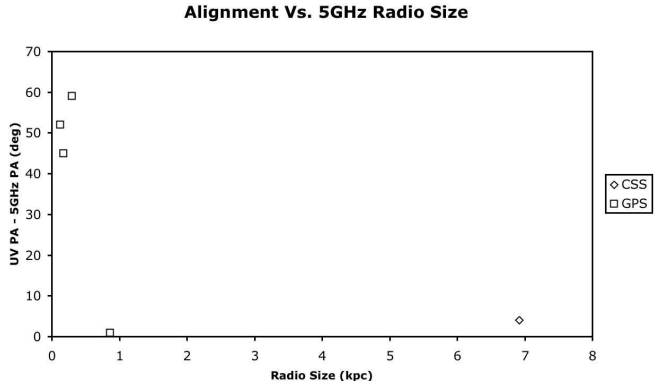}
\caption{Difference between the UV and radio position angles (alignment) of the GPS and CSS galaxies versus radio size.\label{ALLRS}}
\end{figure}

Inspection of Figure \ref{ALLRS} shows that alignment between the UV and radio is found for
sources $\gtrsim 1$ kpc, i.e, the compact GPS sources tend not to show systematic alignment.
If the UV light was scattered nuclear light, it would be expected to align with the radio
axis regardless of radio source size (assuming the scattered light would escape along the radio
axis). The lack of alignment in the GPS sources suggests
that the UV light is not scattered nuclear light. Instead, in the GPS sources, we may
be detecting some clumps of star formation on scales larger than the radio source 
which are associated with the fueling of the radio activity.

\subsection{CSS sources}

The fact that the GPS sources show UV emission on larger scales than the radio source and which are
not aligned with the radio source is consistent with the hypothesis that the GPS sources are
too small to strongly affect their environment on the scales resolved by these observations.
In large scale 3CR sources, the jets extend far beyond the host galaxy.  Therefore, if there is 
jet induced star formation, CSS sources are the best candidates to reveal it. We also note that the two
most luminous UV sources in our sample (\object{1443+77} and \object{1814--637}) are CSS.

\citet{Labiano05} demonstrated the presence of gas ionized by the shocks from the expanding radio source in CSS sources. Furthermore, they found that \object{1443+77} shows the strongest contribution from shocks. It is possible
these shocks are affecting the star formation in the host.

\subsection{Is there jet-induced star formation?}

\begin{figure}[t]
\centering
\begin{center}
\includegraphics[width=\columnwidth]{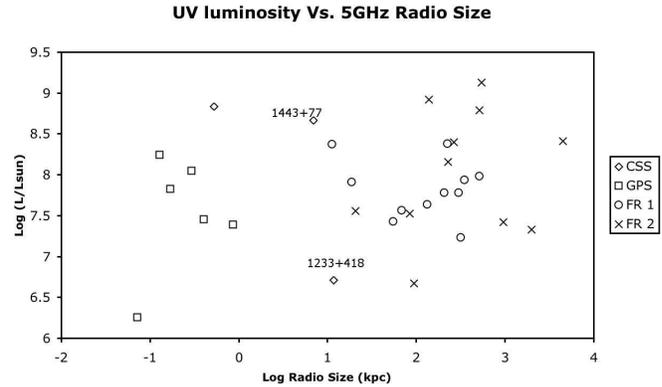}
\caption{As Figure \ref{UVRP}, for radio size.\label{UVRS}}
\end{center}
\end{figure}

Figure \ref{UVRS}  suggests there is no correlation between UV luminosity and size of the radio source
(i.e., the GPS, CSS and large 3CR sources have similar UV luminosities).
Therefore, any UV emission which is related to source size does not seem to dominate the UV
properties of the host. However, if the expansion of the radio lobes were enhancing the UV luminosity,
it could have been unnoticed in our sample. The lack of a significant number of  CSS sources
in our sample (i.e. radio sizes comparable to the host) may cause us to be missing hosts
with UV emission $\sim 10^9$ L$_{\odot}$.

\begin{figure}[h]
\centering
\begin{center}
\includegraphics[width=\columnwidth]{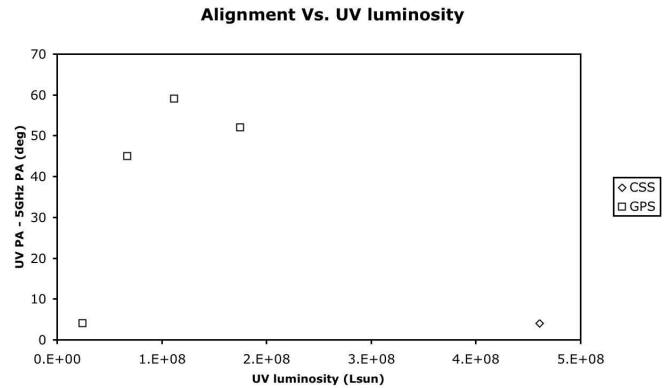}
\caption{As Figure \ref{ALLRS}, for UV luminosity.\label{ALLum}}
\end{center}
\end{figure}

\begin{figure}[h]
\centering
\includegraphics[width=\columnwidth]{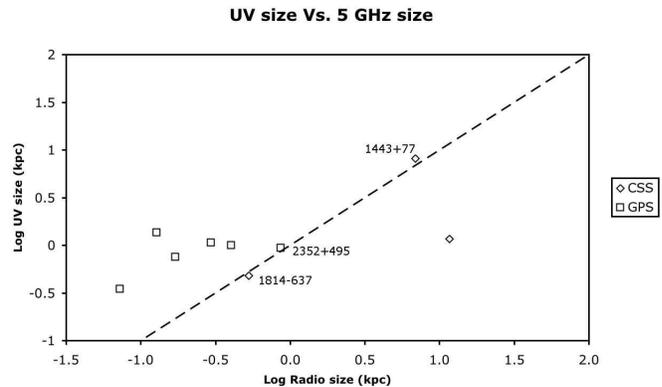}
\caption{UV size and radio size of GPS and CSS sources. The dashed line marks the locus of sources with equal radio and UV sizes.\label{UVSRS}}
\end{figure}

If the jet is enhancing the star formation, we expect the radio source and UV emission to be aligned and have similar sizes. The high UV luminosity source \object{1443+77} shows alignment between the UV and radio source (Figure \ref{ALLum}).
Observations of additional sources with sizes $\gtrsim$ 1 kpc are needed to improve the statistics.
Figure \ref{UVSRS} shows that UV size and radio size are not correlated for most sources.  However,
two CSS and one GPS source (\object{1443+77}, \object{1814--637} and \object{2352+495}) have similar
UV and radio sizes. The radio source in \object{1233+418} is larger than the (marginal) UV emission.
However, the UV magnitudes for \object{1233+418} are close to the detection limit so we may
be missing extended and fainter UV emission in this object.

\begin{figure}[h]
\centering
\includegraphics[width=\columnwidth]{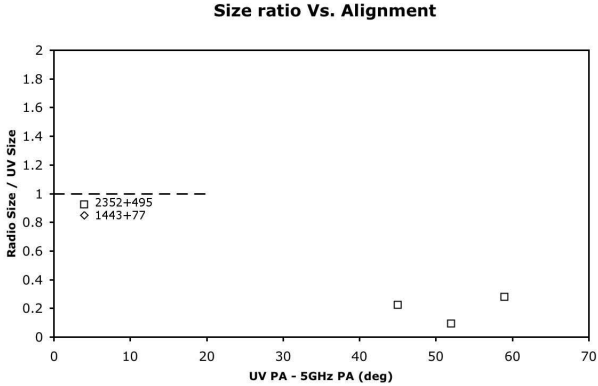}
\caption{Ratio of radio and UV sizes versus the alignment of the GPS and CSS sources. The dashed line marks the locus for sources with same UV and radio size and highest alignment i.e. where the radio source is most likely affecting the star formation in the host. \label{SRAL}}
\end{figure}

Table \ref{uvgraphs} shows the position angles of the radio and UV in the GPS and CSS sources. We see that the CSS \object{1443+77} and the GPS \object{2352+495}  \footnote{For sources with several UV position angles, we use the most aligned component. One of the components aligned with the radio source in \object{2352+495} is a marginal detection.} are aligned with the UV. In addition \object{1443+77} \object{1814--637} and \object{2352+495} have a ratio of radio to UV size of order unity (Figure \ref{SRAL}). The UV emission in \object{2352+495} has two components aligned with the radio source (and a third which is not aligned), but the unknown offset between radio and HST reference frames prevents us from accurately overlaying them. Therefore, \object{1443+77} is currently the best candidate for jet induced star formation.

This is consistent with the hypothesis that the UV emission is dominated by extended regions
of star formation rather than by point-like AGN.



\begin{figure}[t]
\centering
\begin{center}
\includegraphics[width=\columnwidth]{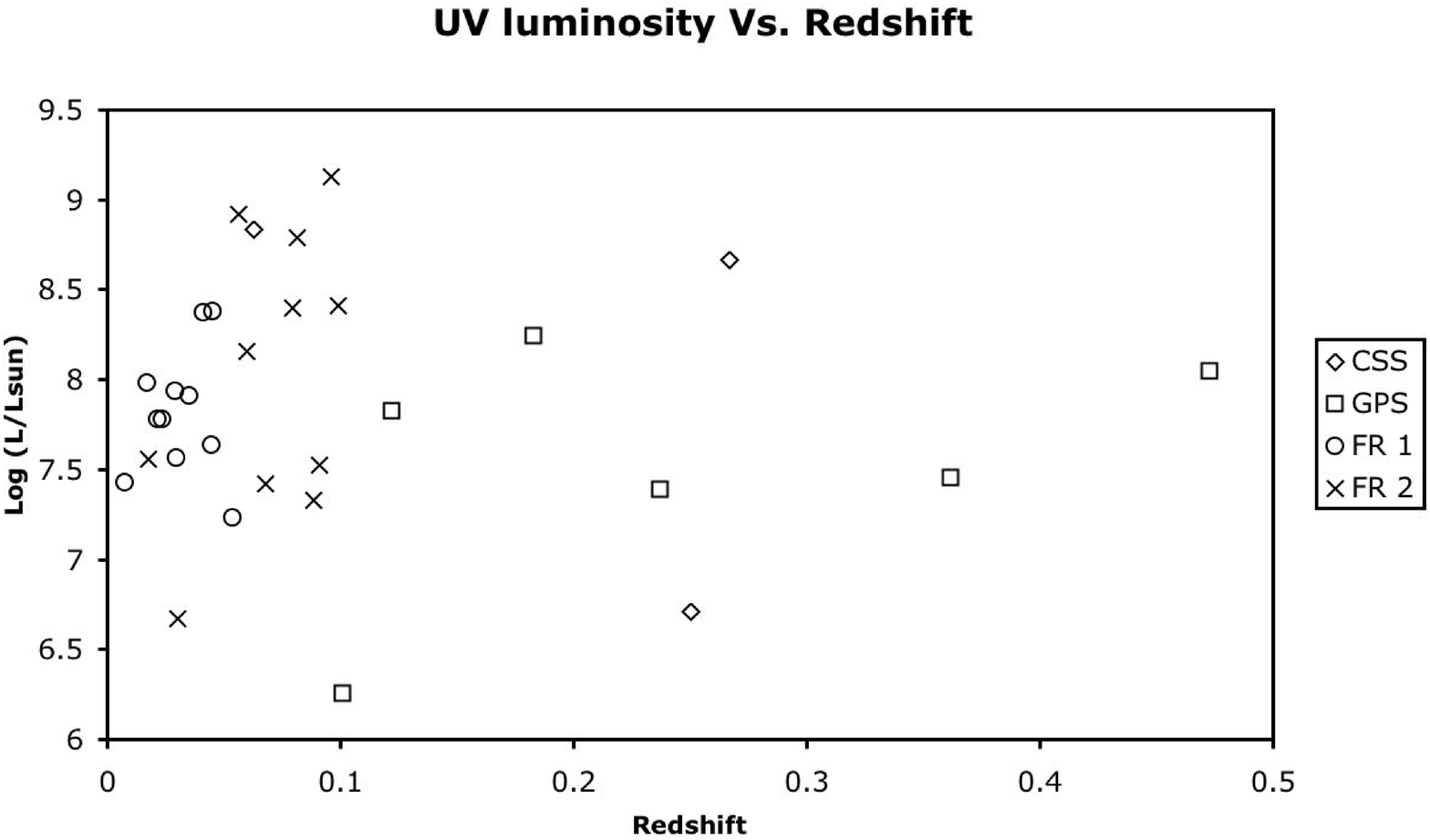}
\caption{As Figure \ref{UVRP}, for redshift. \label{UVZ}}
\end{center}
\end{figure}

\begin{figure}[t]
\centering
\begin{center}
\includegraphics[width=\columnwidth]{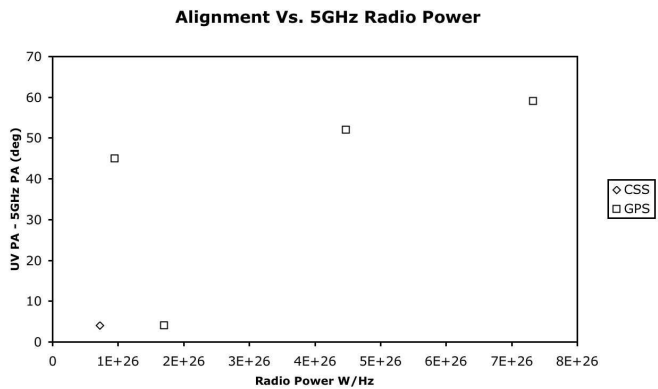}
\caption{As Figure \ref{ALLRS}, for radio power. \label{ALRP}}
\end{center}
\end{figure}

\begin{figure}[h]
\centering
\begin{center}
\includegraphics[width=\columnwidth]{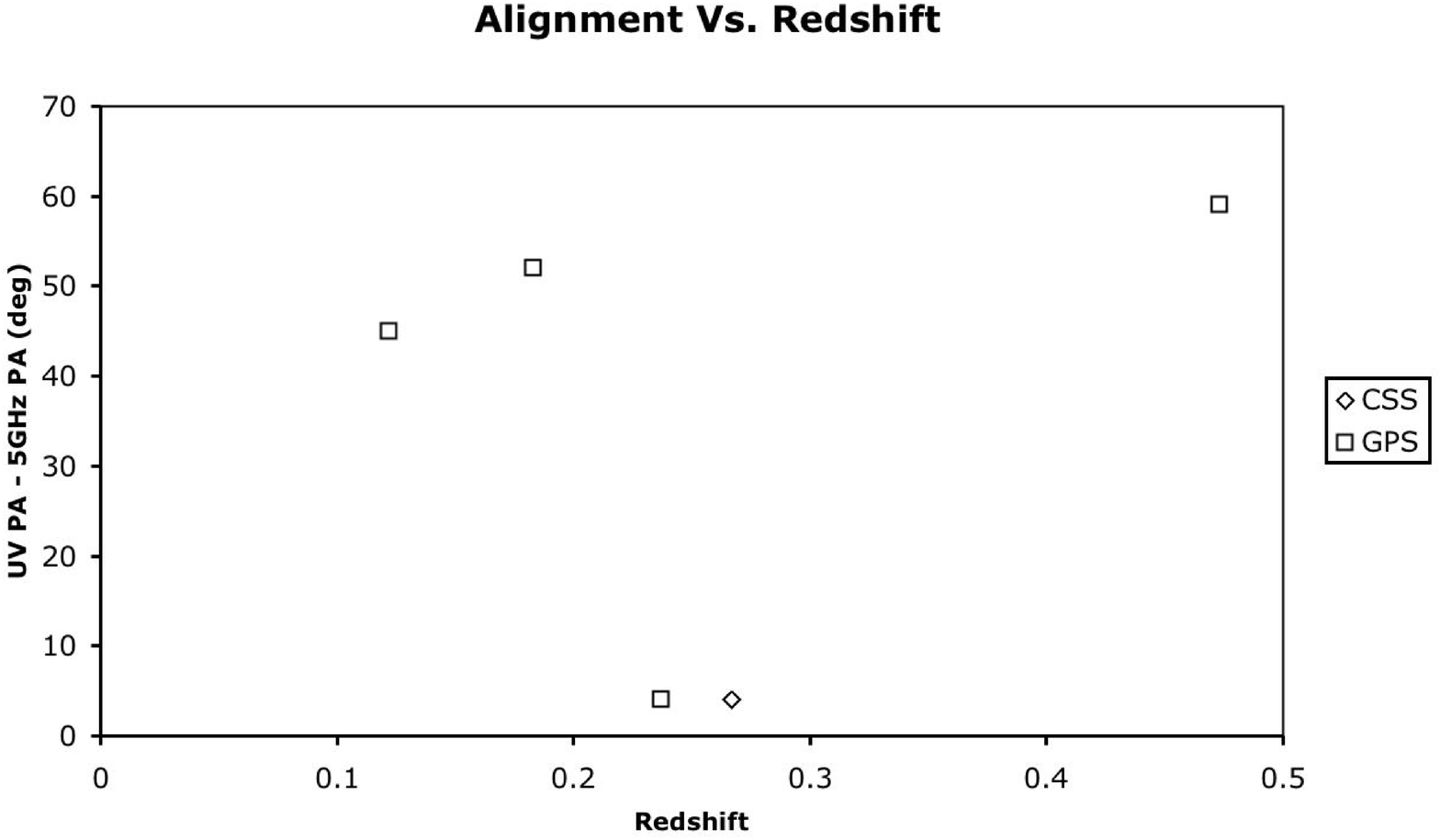}
\caption{As Figure \ref{ALLRS}, for redshift. \label{ALZ}}
\end{center}
\end{figure}

We also note several interesting null results. The detected UV  luminosity in the GPS, CSS,
and large 3CR sources is independent of redshift (out to z=0.5) (Figure \ref{UVZ}).
In the GPS and CSS sources, the alignment does not depend on either radio power (Figure \ref{ALRP})
or redshift (Figure \ref{ALZ}); though in both cases the statistics are dominated by the GPS
sources which seem not to show alignment on the scales resolved by these observations.

\begin{table*}[t]
\caption{Mass from luminosity.} 
\label{msol}
\begin{minipage}{2\columnwidth}
\centering
\begin{tabular}{ccccccc}
\hline
\hline
  &  \multicolumn{2}{c}{GALAXEV CB} & \multicolumn{2}{c}{GALAXEV IB} & \multicolumn{2}{c}{PEGASE} \\ 
Name & 10 Gyr & 1 Gyr & 10 Gyr & 1 Gyr & 10 Gyr & 1 Gyr \\ 
\hline 
\object{1117+146}& 2.2 $\times 10^{11}$& 2.1 $\times 10^{08}$& 2.3 $\times 10^{11}$& 5.5 $\times 10^{09}$& 2.6 $\times 10^{10}$& 8.2 $\times 10^{08}$\\ 
\object{1233+418}& 5.2 $\times 10^{10}$& 1.0 $\times 10^{08}$& 5.6 $\times 10^{10}$& 1.8 $\times 10^{09}$& 1.1 $\times 10^{10}$& 3.9 $\times 10^{08}$\\ 
\object{1345+125}& 1.2 $\times 10^{11}$& 6.7 $\times 10^{08}$& 1.2 $\times 10^{11}$& 6.0 $\times 10^{09}$& 4.7 $\times 10^{10}$& 2.2 $\times 10^{09}$\\ 
\object{1443+77} & 2.0 $\times 10^{12}$& 3.3 $\times 10^{09}$& 2.2 $\times 10^{12}$& 6.2 $\times 10^{10}$& 3.7 $\times 10^{11}$& 1.3 $\times 10^{10}$\\ 
\object{1607+268}& 1.9 $\times 10^{12}$& 1.1 $\times 10^{09}$& 1.9 $\times 10^{12}$& 3.2 $\times 10^{10}$& 1.5 $\times 10^{11}$& 4.3 $\times 10^{09}$\\ 
\object{1814--63} & 1.1 $\times 10^{13}$& 4.6 $\times 10^{09}$& 1.1 $\times 10^{13}$& 2.0 $\times 10^{11}$& 7.1 $\times 10^{11}$& 2.0 $\times 10^{10}$\\ 
\object{1934--638}& 3.0 $\times 10^{11}$& 1.1 $\times 10^{09}$& 3.3 $\times 10^{11}$& 1.3 $\times 10^{10}$& 9.5 $\times 10^{10}$& 3.9 $\times 10^{09}$\\ 
\object{1946+708}& 1.4 $\times 10^{09}$& 9.6 $\times 10^{06}$& 1.5 $\times 10^{09}$& 7.8 $\times 10^{07}$& 6.2 $\times 10^{08}$& 3.1 $\times 10^{07}$\\ 
\object{2352+495}& 4.1 $\times 10^{10}$& 9.0 $\times 10^{07}$& 4.4 $\times 10^{10}$& 1.5 $\times 10^{09}$& 9.3 $\times 10^{09}$& 3.4 $\times 10^{08}$\\ 
\hline
  & 10 Myr & 1 Myr & 10 Myr & 1 Myr & 10 Myr & 1 Myr\\ 
\hline
\object{1117+146}&1.7 $\times 10^{09}$& 1.8 $\times 10^{08}$& 1.4 $\times 10^{06}$& 4.4 $\times 10^{06}$& 5.9 $\times 10^{07}$& 2.5 $\times 10^{08}$\\ 
\object{1233+418} &1.0 $\times 10^{09}$& 1.0 $\times 10^{08}$& 8.3 $\times 10^{05}$& 2.5 $\times 10^{06}$& 3.7 $\times 10^{07}$& 1.8 $\times 10^{08}$\\ 
\object{1345+125} &8.9 $\times 10^{09}$& 8.2 $\times 10^{08}$& 7.1 $\times 10^{06}$& 2.0 $\times 10^{07}$& 3.0 $\times 10^{08}$& 1.4 $\times 10^{09}$\\ 
\object{1443+77}  &3.2 $\times 10^{10}$& 3.1 $\times 10^{09}$& 2.5 $\times 10^{07}$& 7.8 $\times 10^{07}$& 1.1 $\times 10^{09}$& 5.4 $\times 10^{09}$\\ 
\object{1607+268} &8.1 $\times 10^{09}$& 8.3 $\times 10^{08}$& 6.4 $\times 10^{06}$& 2.2 $\times 10^{07}$& 2.8 $\times 10^{08}$& 1.2 $\times 10^{09}$\\ 
\object{1814--63}  &3.2 $\times 10^{10}$& 3.5 $\times 10^{09}$& 2.6 $\times 10^{07}$& 9.5 $\times 10^{07}$& 1.3 $\times 10^{09}$& 6.1 $\times 10^{09}$\\ 
\object{1934--638} &1.3 $\times 10^{10}$& 1.2 $\times 10^{09}$& 1.0 $\times 10^{07}$& 3.0 $\times 10^{07}$& 4.4 $\times 10^{08}$& 2.1 $\times 10^{09}$\\ 
\object{1946+708} &1.3 $\times 10^{08}$& 1.2 $\times 10^{07}$& 1.1 $\times 10^{05}$& 3.0 $\times 10^{05}$& 4.4 $\times 10^{06}$& 2.1 $\times 10^{07}$\\ 
\object{2352+495} &9.4 $\times 10^{08}$& 9.1 $\times 10^{07}$& 7.5 $\times 10^{05}$& 2.3 $\times 10^{06}$& 3.3 $\times 10^{07}$& 1.6 $\times 10^{08}$\\ 
\hline
\end{tabular}
\end{minipage}
\\ \\
{\bf Mass in M$_\odot$ needed to reproduce the observed F330W luminosity, according to the mass to luminosity ratio from each model. CB is the 1 Myr long continuous burst, IB is the instantaneous single burst model, PEGASE is the elliptical template model. The table is divided two halves corresponding to models of Gyr and Myr}.
\end{table*}

\section{Stellar synthesis models}
\label{ssm}

{\bf GPS and CSS sources are usually associated with massive ellipticals of ages $\sim 5$ Gyr and solar metallicities \citep[e.g.,][]{Vries07, Vries00, Vries98b, Snellen98, Snellen99}. With this assumptions, we use stellar population synthesis models by \citet{Bruzual03} using the Chabrier \citep{Chabrier03} initial mass function and Padova evolutionary tracks, for a 1 Myr long continuous burst of star formation and a instantaneous single burst model for populations of 10$^4, 10^3$, 10 and 1 Myr to compare our UV luminosity measurements and to estimate the mass and age of stars producing the UV emission in our sources. We also compare the measurements with the elliptical galaxy templates from PEGASE.}

Table \ref{msol} shows the mass of stars for different ages needed to reproduce the observed near-UV emission of each source. Our observations are generally consistent with models
of a single instantaneous burst of 10$^6 - 10^7$ M$\odot$, $\lesssim$10 Myr ago. However, some
sources may be dominated by intermediate-age (0.1 to 1 Gyr) populations \citep[with masses
$\sim10^9$M$\odot$, e.g.,][]{Tadhunter05}. These young and intermediate ages in our
GPS/CSS sources are also consistent with stellar population ages measured in hosts of powerful
radio galaxies \citep{Raimann05} and other CSS sources \citep[e.g.,][]{Johnston05}.
The models which require old and massive stellar populations to produce the UV light are inconsistent
with the small measured sizes for the UV emitting regions.

Our observations are consistent with a scenario in which a single event is responsible for triggering
the AGN and initiating a starburst. However, the dynamical ages of GPS and CSS sources are much
smaller (between 10$^3$ and 10$^6$ \citep[e.g.,][]{Polatidis03, O'Dea98}) than the estimated ages of
the stellar populations produced in the starburst. This implies
a minimum time delay between the start of the starburst and the start of the radio activity of 10 Myr.
The delay could be as long as 1 Gyr if the 1 Gyr stellar population models are correct.
The apparent delay between starburst and formation of the radio source has been found in objects
with possible connections between the starburst and the formation of the AGN and radio source
\citep{Raimann05, Tadhunter05} and is predicted by theoretical calculations of the time needed
for the gas to reach the center of the galaxy after a tidal interaction \citep[e.g.,][]{Lin88}.



\section{Summary}

We have obtained HST/ACS/HRC near-UV high resolution images of compact GPS and CSS radio galaxies
which are likely to be the young progenitors of large scale powerful radio galaxies.
We detect near-UV emission in point sources and/or small clumps (tens to hundreds of pc
scale) in seven of the sources, consistent with the presence of recent star formation. Two sources show very weak near-UV emission and may have not been detected. 
The UV luminosity in the GPS and CSS sources is similar to that measured in nearby large
3CR radio galaxies by \cite{Allen02}.
In the GPS sources as a whole we do not see systematic alignment between the radio and UV.
This may be because the GPS radio sources are smaller than the scales resolved by these
HST observations. The lack of alignment in the GPS sources implies that the UV emission is
not due to scattered nuclear light.

There is evidence for a correlation between radio and UV luminosity in the GPS sources.
This relationship could be produced by having  a larger reservoir
of gas available in the more luminous sources
for fueling the radio source and supplying the starburst.

In the CSS source \object{1443+77}  the near-UV emission is aligned with and co-spatial with the the
radio emission and we suggest that star formation has been triggered or at least enhanced
by expansion of the radio source through the host.

We suggest that the starburst and AGN are triggered by the same event.
Comparison with stellar population synthesis models suggests the data are consistent with
star formation occurring in a burst $\lesssim$10 Myr ago (though models as old as 1 Gyr are
not ruled out). The radio source ages are much smaller
(e.g., $10^3$ to $10^6$ yr) suggesting that there is a minimum delay of 10 Myr between the onset
of the starburst and the onset of the radio activity \citep[see also][]{Raimann05, Tadhunter05}.
Observations at other wavelengths and measurement of the colors are needed to further  
asses the nature of the observed UV properties.


\begin{acknowledgements}
We thank the anonymous referee for valuable suggestions and comments on the paper. AL wishes to thank Dr. Peng (STScI) for his help with GALFIT and Dr. I.A.G. Snellen (Leiden) for fruitful scientific discussions. WDVs work was performed under the auspices of the U.S. Department of Energy, National Nuclear Security Administration by the University of California, Lawrence Livermore National Laboratory under contract No. W-7405-Eng-48. These observations are associated with program 10117. Support for program 10117 was provided by NASA through a grant from the Space Telescope Science Institute. This research has made use of NASA's Astrophysics Data System Bibliographic Services and of the NASA/IPAC Extragalactic Database (NED) which is operated by the Jet Propulsion Laboratory, California Institute of Technology, under contract with the National Aeronautics and Space Administration. This research is based on observations made with the NASA/ESA Hubble Space Telescope, obtained fro
 m the data archive at the Space Telescope Institute. STScI is operated by the association of Universities for Research in Astronomy, Inc. under the NASA contract  NAS 5-26555.
\end{acknowledgements}


\end{document}